\documentclass[prd,aps,twocolumn,superscriptaddress,showpacs,showkeys,amsmath,amssymb,floatfix]{revtex4}
\usepackage[colorlinks=true,citecolor=blue,filecolor=blue,linkcolor=blue,urlcolor=blue,pdftex]{hyperref}

\usepackage{xspace} 
\usepackage[usenames,dvipsnames]{color}
\usepackage{booktabs,graphicx,mathrsfs,verbatim,amsmath,units,soul}
\usepackage{xspace} 
\usepackage{upgreek}
\usepackage{gensymb}
\usepackage{amsmath}
\usepackage{tabularx}
\usepackage[none]{hyphenat}
\usepackage{multirow}
\usepackage{textcomp}
\usepackage [autostyle, english = american]{csquotes}
\MakeOuterQuote{"}

\newcommand{\SRzero}{SR0\xspace}
\newcommand{\SRone}{SR1\xspace}

\newcommand{\rnzero}{${}^{220}$Rn\xspace}

\newcommand{\cSone}{\ensuremath{\mathrm{cS1}}\xspace}
\newcommand{\cStwo}{\ensuremath{\mathrm{cS2}}\xspace}
\newcommand{\cStwob}{\ensuremath{\mathrm{cS2_b}}\xspace}

\newcommand{\radius}{R\xspace}

\newcommand{\radiusm}{\ensuremath{\mathrm{R}}}

\newcommand{\micro}{\ensuremath{\textmu}}

\newcommand{\gevcsq}{\ensuremath{\mathrm{GeV}/\mathrm{c}{}^2\xspace}}



\newcommand{\bologna}{\affiliation{Department of Physics and Astronomy, University of Bologna and INFN-Bologna, 40126 Bologna, Italy}}

\newcommand{\chicago}{\affiliation{Department of Physics \& Kavli Institute for Cosmological Physics, University of Chicago, Chicago, IL 60637, USA}}

\newcommand{\coimbra}{\affiliation{LIBPhys, Department of Physics, University of Coimbra, 3004-516 Coimbra, Portugal}}

\newcommand{\columbia}{\affiliation{Physics Department, Columbia University, New York, NY 10027, USA}}

\newcommand{\lngs}{\affiliation{INFN-Laboratori Nazionali del Gran Sasso and Gran Sasso Science Institute, 67100 L'Aquila, Italy}}

\newcommand{\mainz}{\affiliation{Institut f\"ur Physik \& Exzellenzcluster PRISMA, Johannes Gutenberg-Universit\"at Mainz, 55099 Mainz, Germany}}

\newcommand{\heidelberg}{\affiliation{Max-Planck-Institut f\"ur Kernphysik, 69117 Heidelberg, Germany}}

\newcommand{\munster}{\affiliation{Institut f\"ur Kernphysik, Westf\"alische Wilhelms-Universit\"at M\"unster, 48149 M\"unster, Germany}}

\newcommand{\nikhef}{\affiliation{Nikhef and the University of Amsterdam, Science Park, 1098XG Amsterdam, Netherlands}}

\newcommand{\nyuad}{\affiliation{New York University Abu Dhabi, Abu Dhabi, United Arab Emirates}}

\newcommand{\purdue}{\affiliation{Department of Physics and Astronomy, Purdue University, West Lafayette, IN 47907, USA}}

\newcommand{\rpi}{\affiliation{Department of Physics, Applied Physics and Astronomy, Rensselaer Polytechnic Institute, Troy, NY 12180, USA}}

\newcommand{\rice}{\affiliation{Department of Physics and Astronomy, Rice University, Houston, TX 77005, USA}}

\newcommand{\stockholm}{\affiliation{Oskar Klein Centre, Department of Physics, Stockholm University, AlbaNova, Stockholm SE-10691, Sweden}}

\newcommand{\subatech}{\affiliation{SUBATECH, IMT Atlantique, CNRS/IN2P3, Universit\'e de Nantes, Nantes 44307, France}}

\newcommand{\torino}{\affiliation{INFN-Torino and Osservatorio Astrofisico di Torino, 10125 Torino, Italy}}

\newcommand{\ucla}{\affiliation{Physics \& Astronomy Department, University of California, Los Angeles, CA 90095, USA}}

\newcommand{\ucsd}{\affiliation{Department of Physics, University of California, San Diego, CA 92093, USA}}

\newcommand{\wis}{\affiliation{Department of Particle Physics and Astrophysics, Weizmann Institute of Science, Rehovot 7610001, Israel}}

\newcommand{\zurich}{\affiliation{Physik-Institut, University of Zurich, 8057  Zurich, Switzerland}}

\newcommand{\paris}{\affiliation{LPNHE, Universit\'{e} Pierre et Marie Curie, Universit\'{e} Paris Diderot, CNRS/IN2P3, Paris 75252, France}}

\newcommand{\freiburg}{\affiliation{Physikalisches Institut, Universit\"at Freiburg, 79104 Freiburg, Germany}}

\newcommand{\lal}{\affiliation{LAL, Universit\'e Paris-Sud, CNRS/IN2P3, Universit\'e Paris-Saclay, F-91405 Orsay, France}}

\newcommand{\naples}{\affiliation{Department of Physics ``Ettore Pancini'', University of Napoli and INFN-Napoli, 80126 Napoli, Italy}} 

\newcommand{\nagoya}{\affiliation{Kobayashi-Maskawa Institute for the Origin of Particles and the Universe, Nagoya University, Furo-cho, Chikusa-ku, Nagoya, Aichi 464-8602, Japan}}

\begin{document}


\title{XENON1T Dark Matter Data Analysis: Signal \& Background Models, and Statistical Inference.}


\author{E.~Aprile}\columbia
\author{J.~Aalbers}\stockholm\nikhef
\author{F.~Agostini}\bologna
\author{M.~Alfonsi}\mainz
\author{L.~Althueser}\munster
\author{F.~D.~Amaro}\coimbra
\author{V.~C.~Antochi}\stockholm
\author{F.~Arneodo}\nyuad
\author{L.~Baudis}\zurich
\author{B.~Bauermeister}\stockholm
% ...
\collaboration{XENON Collaboration}
\email[]{xenon@lngs.infn.it}
\noaffiliation

\date{\today} 

\begin{abstract}


The XENON1T experiment searches for dark matter particles through their scattering off xenon atoms in a 2\,tonne liquid xenon target.
The detector is a dual-phase time projection chamber, which measures simultaneously the scintillation and ionization signals produced by interactions in target volume, to reconstruct energy and position, as well as the type of the interaction. 
The background rate in the central volume of XENON1T detector is the lowest achieved so far with a liquid xenon-based direct detection experiment.
In this work we describe the response model of the detector,  the background and signal models, and the statistical inference procedures used in the dark matter searches with a 1\,tonne$\times$year exposure of XENON1T data, that leaded to the best limit to date on WIMP-nucleon spin-independent elastic scatter cross-section for WIMP masses above 6\,GeV/c$^2$.
\end{abstract}



\maketitle
\section{Introduction}

The existence of dark matter (DM) and its making up about 26\% of the mass-energy of the Universe~\cite{aghanim2018planck} is indicated by a wide range of astronomical and cosmological observations.
Direct detection experiments, which search for DM particles interacting with ordinary matters in a terrestrial detector target, have not yet yielded unequivocal evidence for dark matter~\cite{akerib2017results,aprile2016xenon100, cui2017dark,abea2018direct,petricca2017first,agnese2015wimp}.
The XENON1T experiment~\cite{aprile2017xenon1t}, located in the INFN Laboratori Nazionali del Gran Sasso, Italy, primarily searches for Weakly Interacting Massive Particles (WIMPs), which could scatter elastically off xenon atoms. 
Using a 1\,tonne$\times$year exposure and a nuclear recoil (NR) energy range from 4.9\,keV to 40.9\,keV, XENON1T has set upper limits on the cross section of spin-independent elastic scattering with a minimum of 4.1$\times$10$^{-47}$ cm$^2$ for a 30\,GeV/c$^2$ WIMP~\cite{xe1t_sr1}. These are the most stringent constraints set on this interaction for WIMP masses above 6\,GeV/c$^2$.
The XENON1T experiment has  achieved the lowest background rate among liquid xenon (LXe) detectors to date.

The dual-phase time projection chamber (TPC)  used by the XENON1T detector allows for the reconstruction of the deposited energy and the three-dimensional position of interactions in the active liquid xenon target. The observable signals are the scintillation (S1) and ionization (S2) signals produced by energy depositions. 
The longitudinal ($z$) position is reconstructed using the time difference between the prompt S1 signal and the S2 signal, which is produced by electroluminescence in gaseous xenon after electrons, drifted upwards by an electric field, get extracted from the liquid into the gas. Both signals are observed by arrays of photomultiplier tubes (PMTs) arranged at the top and bottom of the detector. 
The position in the ($x$,$y$) plane is reconstructed using the S2 signal pattern in the upper PMT array.
Background from radioactivity in detector materials can be rejected to a large extent by selecting a three-dimensional fiducial region within the active volume. 
In addition, the S2-S1 ratio can be used to discriminate between NRs from WIMPs and neutrons and electronic recoils (ERs) from $\gamma$ and $\beta$, which constitute the major backgrounds of the XENON1T experiment.
More details on TPC working principles and the XENON1T TPC can be found in~\cite{aprile2017xenon1t}.

The XENON1T data analysis can be divided into two parts. 
The first part includes event reconstruction, signal corrections, and event selection, and is reported in~\cite{xe1t_long_analysis_1}. The second part includes the detector response model, the background and WIMP signal models, and the statistical inference, and is presented in detail in this manuscript.
These models and techniques are used in the XENON1T DM searches~\cite{xe1t_sr0, xe1t_sr1}.
The detector response model, which will be presented in Section~\ref{sec:detector_signal_response_model}, describes how an ER or a NR energy deposition is reconstructed in the TPC. The fit of the detector response model to calibration data provides the ER and NR background models and the signal model described in Section~\ref{sec:background_models}, which also considers background models constructed using data-driven methods. Lastly, the statistical inference is presented in Section~\ref{sec:statistical_inference}.
A summary is then given in Section~\ref{sec:summary_outlook}.

\section{Detector Response Model}

\label{sec:detector_signal_response_model}

\label{sec:signal_response_model}
Understanding the conversion from the deposited energy to the observed S1 and S2 signals is critical for interpreting the results of DM searches in XENON1T.
The energy region of interest is in the range of a few keV to several tens of keV in searches for elastic scatters between WIMP and xenon nuclei.
The conversion of deposited energy to S1 and S2 in this region is non-linear, with fluctuations due to the scintillation and ionization processes in LXe and due to the detector reconstruction.  

The model of the XENON1T detector response to ERs and NRs is based on simulations which include a comprehensive description of the signal production process and detailed characterizations of detector detection and reconstruction effects. The model is constrained by a Bayesian simultaneous fit to ER and NR calibration data, which allows to use all available information and treat correlated detector uncertainties coherently.

\subsection{Basic Signal Response in Liquid Xenon}

The intrinsic signal response model in LXe follows the approach used in the Noble Element Simulation Technique (NEST) model~\cite{nest_v0.98,nest_v1.0}.
There are three forms of energy deposition in LXe: thermalization of the recoiling particle, excitation of xenon atoms, and ionization of xenon atoms. The thermalization energy loss is undetectable in the XENON1T detector. 
The number of detectable quanta $N_q$ is the sum of the number of excitons $N_{\mathrm{ex}}$ and ion-electron pairs $N_i$, and can be used to reconstruct the deposited energy $\varepsilon$.
It follows a Binomial fluctuation due to the potential energy loss to thermalization,

\begin{equation}
\large
N_q \sim \mathrm{Binom}\left( \varepsilon/W, L\right),
    \label{eqn:fluc_number_of_quanta}
\end{equation}
where $L$ is the Lindhard factor expressing the fraction of energy loss to heat and $W$ (13.7$\pm$0.2\,eV from a global fit~\cite{nest_v0.98}) is the average energy required to create either an exciton or ion-electron pair in LXe. 
Negligible energy is lost to thermalization in an ER as the mass of the recoiling electron is much smaller than the xenon nucleus.
In NRs, the recoiling xenon atom transfers kinetic energy through elastic scattering off surrounding xenon atoms, resulting in a Lindhard factor~\cite{lindhard1963integral} of 0.1-0.2 in LXe. The field- and energy-dependence of the Lindhard factor are parametrized following the NEST model~\cite{nest_v1.0}.

The exciton-to-ion ratio $\langle N_{\mathrm{ex}}/N_i \rangle$ is related to the excitation and ionization cross sections of recoiling particles on xenon atoms.
For ERs, it is assumed to be constant, and is given a uniform prior ranging from 0.06 to 0.20~\cite{nest_v0.98}.
For NRs, it is parametrized as a function of the deposited energy and the electric field strength $F$ in the active volume, following~\cite{nest_v1.0}, and is in the range of 0.7 to 1.0 for NR energies from 5 to 40\,keV$_{\mathrm{nr}}$ under a field of 81\,kV/cm.
The binomial fluctuations of $N_{ex}$ and $N_i$ can be written as

\begin{equation}\large
\begin{array}{rcl}
N_i & \sim & \mathrm{Binom} \left( N_q, \frac{1}{1+\langle N_{\mathrm{ex}}/N_i \rangle} \right), \\
N_{\mathrm{e}x} & = & N_q - N_i.
\end{array}
    \label{eqn:fluc_excimer_to_ion_ratio}
\end{equation}
Ionization electrons have a probability $1-r$ to escape the cloud of ion-electron pairs, where $r$ is referred to as the recombination fraction,

\begin{equation}\large
\begin{array}{rcl}
N_e & \sim & \mathrm{Binom} \left( N_i, 1-r \right), \\
N_{\gamma} & =  & N_i - N_e + N_{\mathrm{ex}},
\end{array}
    \label{eqn:recombination_process}
\end{equation}
where $N_{\gamma}$ and $N_e$ are the number of photons generated by de-excitation of the initial excitons and ion-electron recombination, and of the escaping electrons, respectively.
Due to detector effects, such as field non-uniformity, and intrinsic fluctuations~\cite{lux_tritium}, the recombination fraction $r$ fluctuates, and is modeled as a Gaussian distribution,

\begin{equation}\large
r \sim \mathrm{Gauss} \left( \langle r \rangle, \Delta r \right).
    \label{eqn:recombination_fluctuation}
\end{equation}
The mean recombination fraction $\langle r \rangle$ depends on the deposited energy and the electric field, and is described by the Thomas-Imel (TI) box model~\cite{ti_box_model},

\begin{equation}\large
\langle r \rangle = 1 - \frac{\ln{(1+N_i\varsigma/4)}}{N_i\varsigma/4},
    \label{eqn:ti_box_model}
\end{equation}
where $\varsigma$ is the field-dependent TI model parameter.
$\Delta r$ is the recombination fluctuation, parameterized as

\begin{equation}\large
   \Delta r = q_2 (1 - e^{-\varepsilon/q_3}),
    \label{eqn:recomb_fluc_parameterization}
\end{equation}
where $q_2$ and $q_3$ are free parameters.
The parametrization is empirically chosen to take into account both the fact that a constant recombination fluctuation is observed with deposit energy larger than 2\,keV and the assumption that the recombination fluctuation diminishes as deposit energy goes to zero.

\begin{figure*}[htp]
    \centering
    \includegraphics[width=0.9\textwidth]{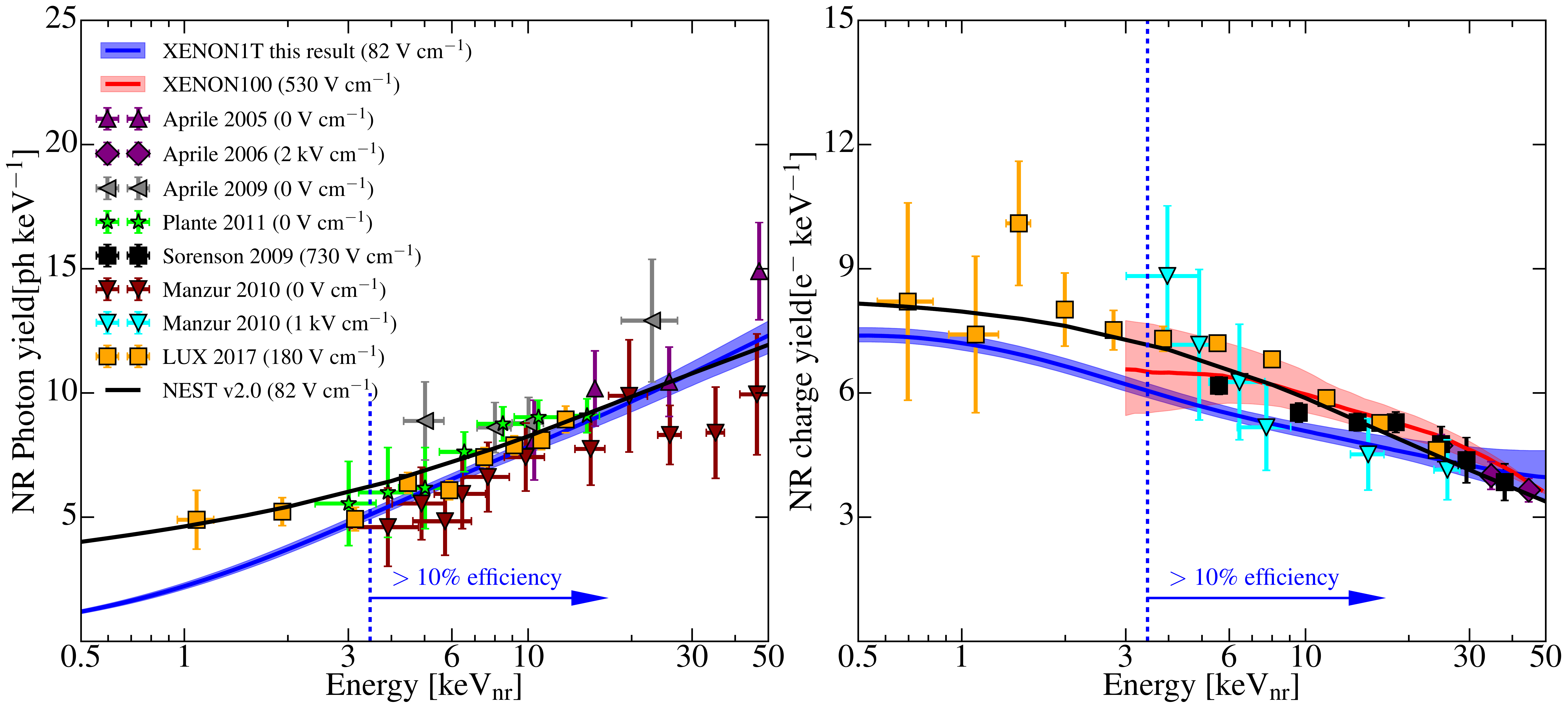}
    \includegraphics[width=0.9\textwidth]{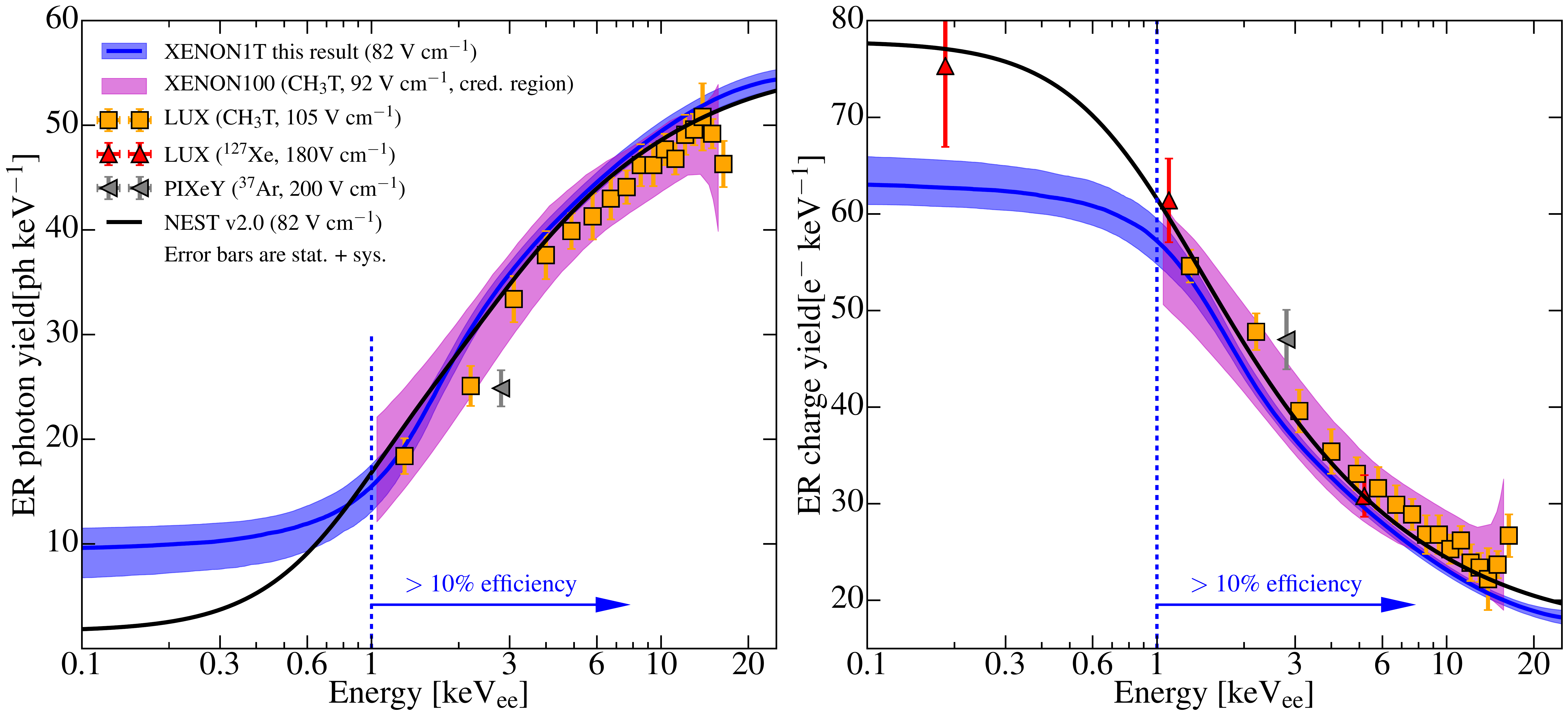}
    \caption{
    Mean photon and charge yields of NR (upper panels) and ER (lower panels) in the XENON1T calibration data fit.
    The blue solid line and shaded region represent the point estimation and 15\%-85\% credible region, respectively, of the posterior.
    Data points for upper panels are from fixed-angle neutron scattering measurements~\cite{aprile2005scintillation,aprile2006simultaneous,aprile2009new,plante2011new,sorensen2009scintillation,manzur2010scintillation,akerib2016low}.
    Results of XENON100~\cite{xenon100_nr} using data-Monte Carlo (MC) matching on the $^{241}$AmBe calibration method are shown with the red solid line and shaded region.
    The best fit from NEST v2.0 ~\cite{nest_v2.0} is shown with the black solid line.
    The measurements from~\cite{tritium_lux,collaboration2018signal,xe127_lux,ar37_lux} are shown along with the best fit of NEST v2.0~\cite{nest_v2.0} in lower panels.
    The vertical dashed blue lines indicates the energy threshold for XENON1T NR and ER calibrations, below which the detection efficiency drops to less than 10\%.
    }
    \label{fig:nr_posterior}
\end{figure*}

For NR, the TI box model, together with the Lindhard factor, has been shown to match data well~\cite{nest_v1.0}.
The recombination fluctuation for NRs was shown to be much smaller than the statistical fluctuations on recombination~\cite{xenon100_nr} induced by Eq.~\ref{eqn:fluc_excimer_to_ion_ratio} and~\ref{eqn:recombination_process}, and is set to 0 in the detector response model. 
For the parameterization of the NR recombination fraction $\langle r \rangle_{\mathrm{nr}}$, the TI parameter $\varsigma_{nr}$ is expressed, following~\cite{nest_v1.0}, as a power law function for the field dependence.
For low energy ERs (roughly above 3\,keV and below 10\,keV), recent measurements~\cite{nerix_er,lux_tritium,xe127_lux,ar37_lux} indicate that the TI box model cannot fully describe the recombination process.
Therefore, we use a modified TI box model for the ER recombination fraction $\langle r \rangle_{\mathrm{er}}$,

\begin{equation}
\large
\langle r \rangle_{er}  =  \left( 1 - \frac{\text{ln}(1+N_i\varsigma_{\mathrm{er}}/4)}{(N_i \varsigma_{\mathrm{er}}/4)} \right) / \left( 1 + e^{-(\varepsilon-q_0)/q_1} \right), 
\label{eqn:er_recomb_parameterization1}
\end{equation}

\begin{equation}
\large
\varsigma_{\mathrm{er}}  =  \gamma_{\mathrm{er}} e^{-\varepsilon/\omega_{\mathrm{er}}} F^{-\delta_{\mathrm{er}}}, 
    \label{eqn:er_recomb_parameterization2}
\end{equation}
where the Fermi-Dirac term in Eq.~\ref{eqn:er_recomb_parameterization1} and the exponential term in Eq.~\ref{eqn:er_recomb_parameterization2} were empirically added to the TI box model to account for the deviation of measurements in the $<$3\,keV and $>$10\,keV energy ranges, respectively.
Similarly to NR, the field dependence of the ER TI box parameter $\varsigma_{\mathrm{er}}$ follows a power law as introduced in NEST~\cite{nest_v1.0}.
The free parameters $q_0$, $q_1$, $\gamma_{\mathrm{er}}$, $\omega_{\mathrm{er}}$, and $\delta_{\mathrm{er}}$ are obtained by matching the detector response model to XENON1T data, without any additional constraints.
It is worth noting that NEST model has been updated based on a global fit using recent measurements~\cite{nest_v2.0}, and is compatible with this work in the energy region of interest.
Fig.~\ref{fig:nr_posterior} shows the mean photon and charge yields as a function of energy for NR and ER, respectively, together with  the measurements from~\cite{aprile2005scintillation,aprile2006simultaneous,aprile2009new,xenon100_nr,plante2011new,sorensen2009scintillation,manzur2010scintillation,akerib2016low} for NR and~\cite{collaboration2018signal,xe127_lux,lux_tritium,ar37_lux} for ER.
The mean photon $\langle N_{\gamma}\rangle /\varepsilon$ and charge yields $\langle N_e \rangle /\varepsilon$ are defined as

\begin{equation}\large
\begin{array}{ccl}
\langle N_{\gamma}\rangle/\varepsilon  & = & \frac{1}{W} \frac{\langle r \rangle + \langle N_{\mathrm{ex}}/N_i \rangle}{1+\langle N_{\mathrm{ex}}/N_i \rangle}, \\
\langle N_e\rangle/\varepsilon  & = & \frac{1}{W} \frac{1 - \langle r \rangle}{1+\langle N_{\mathrm{ex}}/N_i \rangle}.
\end{array}
    \label{eqn:photon_charge_yields_definition}
\end{equation}

The calibration of low energy ERs in XENON1T is performed using an internal $^{220}$Rn source~\cite{aprile2017first}.
The energy spectrum of $\beta$-decays from $^{212}$Pb, one of the progenies of $^{220}$Rn, is similar to the dominant ER background, from $\beta$-decays of  $^{214}$Pb originating from $^{222}$Rn emanation, in the low energy region ($<$10\,keV).
However, the detector response model built for ERs in XENON1T is, in principle, not applicable to $\gamma$-induced ERs that at sufficiently high energies may interact with the inner-shell electrons.
When this happens, the vacancy in the inner shell results in either X-ray  or Auger electrons emission, both of which further ionize xenon atoms.
Consequently, $\gamma$-induced ERs can have multiple recoiling electrons instead of one as in $\beta$-induced ERs.
The binding energy for L-shell electron in xenon is about 4.8-5.5\,keV.
According to the NIST database~\cite{wagner2003nist}, the corresponding X-ray has mean free path of about 5\,$\micro$m.
The effect of the separation of electron clouds at this spatial scale on the recombination is not yet understood.

\subsection{Detector Reconstruction Effects}

Besides the intrinsic response of LXe, detector reconstruction effects on the S1 and S2 signals are modeled.
More specifically, the spatial dependence of S1 and S2 signals, the single and double photoelectron (PE) emission of the PMT photocathode~\cite{faham2015measurements,paredes2018response}, the position reconstruction uncertainty, the reconstruction efficiency, bias, and signal fluctuations, and the acceptance of data selections in analysis are taken into account in the model.

Photons from an energy deposition and the subsequent recombination (Eq.~\ref{eqn:recombination_process}) are detected by the PMTs as an S1 with an efficiency, which is the product of the light collection efficiency $\epsilon_L$, PMTs' average quantum efficiency $\epsilon_{\mathrm{QE}}$, and PMTs' average collection efficiency $\epsilon_{\mathrm{CE}}$. Electrons are drifted to the gas-liquid interface under the drift field, and are extracted under the stronger field, amplifying the electron signal (S2) by the gas gain $G$, which is the number of photoelectrons per electron that is extracted into gaseous xenon.
Both $\epsilon_L$ and $G$ are spatially dependent, and related to the energy scale parameters $g^{\prime}_1$ (probability of one emitted photon to be detected as one PE) and $g^{\prime}_2$ (amplification factor for charge signal), respectively, by

\begin{equation}
    \large
    \begin{array}{ccl}
    g^{\prime}_1 (x, y, z) & = & (1+p_{\mathrm{dpe}}) \cdot \epsilon_L (x, y, z) \cdot \epsilon_{\mathrm{QE}} \cdot \epsilon_{\mathrm{CE}}, \\
    g^{\prime}_2 (x,y) & = & \epsilon_{\mathrm{ext}} G (x,y),
    \end{array}
    \label{eqn:g1_g2_connection}
\end{equation}
where $p_{\mathrm{dpe}}$ is the probability for the PMT photocathode to emit two photoelectrons when absorbing one photon~\cite{faham2015measurements,paredes2018response}, and $\epsilon_{\mathrm{ext}}$ is the extraction efficiency of the drifted electrons which is assumed to be constant in this study.
Note that $g_1$ and $g_2$ in~\cite{xe1t_long_analysis_1} correspond to the averages of $g^{\prime}_1 (x, y, z)$ and $g^{\prime}_2 (x,y)$, respectively, in Eq.~\ref{eqn:g1_g2_connection} over the active volume.
The number of hits detected by PMTs, $N_{\mathrm{hit}}$, and photoelectrons generated from the PMT photocathode, $N_{\mathrm{pe}}$, can be described by a binomial distribution

\begin{equation}\large
    \begin{array}{ccl}
    N_{hit} & \sim & \text{Binom}\left( N_{\gamma},  \epsilon_L (x, y, z) \cdot \epsilon_{\mathrm{QE}} \cdot \epsilon_{\mathrm{CE}} \right), \\
    N_{\mathrm{pe}} - N_{\mathrm{hit}} & \sim & \text{Binom} \left( N_{\mathrm{hit}}, p_{\mathrm{dpe}} \right).
    \end{array}
    \label{eqn:s1_signal_detection}
\end{equation}

In addition to the ($x$, $y$) dependence caused by the varying charge amplification, S2 signals are a function of the $z$ position, because the electrons attach to electronegative impurities when drifting towards the gaseous phase.
The number of electrons $N_{\mathrm{ext}}$ that survive the drifting and the extraction into the gas can be modeled as

\begin{equation}\large
N_{\mathrm{ext}} \sim \text{Binom} \left( N_e, e^{-z/(\tau_e \cdot \nu_d)} \epsilon_{\mathrm{ext}} \right),
    \label{eqn:number_of_extracted_electron}
\end{equation}
where $\tau_e$ and $\nu_d$ are the electron lifetime and electron drift velocity, respectively.
The total proportional scintillation light detected, $N_{\mathrm{prop}}$, can be approximated as

\begin{equation}\large
N_{\mathrm{prop}} \sim \text{Gauss}\left( N_{\mathrm{ext}} G, \sqrt{N_{\mathrm{ext}}} \Delta G \right),
    \label{eqn:number_of_proportional_light}
\end{equation}
where $\Delta G$ is the spread of the gas gain. 
For simplicity, we consider $\Delta G/G$ a constant in the model.

The S1 and S2 signals are constructed from $N_{\mathrm{pe}}$ and $N_{\mathrm{prop}}$, respectively, amplified by the PMTs, digitized, and selected by clustering and classification software~\cite{pax}. To account for biases and fluctuations in this process, the  S1 and S2 are written as

\begin{equation}\large
    \begin{array}{ccl}
         \mathrm{S1}/N_{\mathrm{pe}} - 1 & \sim & \text{Gauss} \left( \delta_{\mathrm{s1}}, \Delta \delta_{\mathrm{s1}} \right), \\
         \mathrm{S2}/N_{\mathrm{prop}} - 1 & \sim & \text{Gauss} \left( \delta_{\mathrm{s2}}, \Delta \delta_{\mathrm{s2}} \right), 
    \end{array}
    \label{eqn:uncorrected_s1_s2}
\end{equation}
where $\delta_{\mathrm{s1}}$ ($\delta_{\mathrm{s2}}$) and $\Delta \delta_{\mathrm{s1}}$ ($\Delta \delta_{\mathrm{s2}}$) are the bias and spread, respectively, of the S1 (S2) reconstruction.
Reconstruction biases and fluctuations are estimated using a waveform simulation including  models of realistic scintillation pulse shape, charge amplification, electronic noise level, PMT single PE spectrum, PMT after-pulses, as well as secondary S2s induced by photonionizations on grids and impurities in the LXe volume~\cite{pax,xe1t_long_analysis_1}.

The S1 and S2 signals are corrected for their spatial dependence based on the reconstructed positions $\vec{x}_{r}$.
The $z$ position of an event is reconstructed using the time difference between the S1 and the S2 signals, and has better resolution than the ($x$, $y$) position which is reconstructed through the S2 hit pattern on the PMTs in the top array.
We assume the reconstruction fluctuations along $x$ and $y$ axes to be identical.
The reconstructed position of $\vec{x}_r$ can be written as

\begin{equation}\large
\vec{x}_r \sim \text{Gauss} \left( \vec{x}, \sigma_p \right),
    \label{eqn:reconstructed_x}
\end{equation}
where $\sigma_p$ is the position reconstruction resolution, and depends on both S2 area and the ($x,y,z$) position of event.
The corrected S1 (cS1) and S2 (cS2) are written as

\begin{equation}\large
    \begin{array}{ccl}
        \text{cS1} & = & \text{S1} \frac{g_1}{g^{\prime}_1 (x_r, y_r, z_r)},  \\
        \text{cS2} & = & \text{S2} \frac{g_2}{g^{\prime}_2 (x_r, y_r)} e^{z/(\tau_e^{\prime} \cdot \nu_d)},
    \end{array}
    \label{eqn:s1_s2_correction}
\end{equation}
where $\tau_e^{\prime}$ is the mean electron lifetime measured. 
The uncertainty of the measured electron lifetime is used to constrain $\tau_e$ in the signal response model.
The correction is based on the measured electron lifetime $\tau_e^{\prime}$.
In the following analysis, as well as in XENON1T results~\cite{xe1t_sr0,xe1t_sr1}, we use the corrected S2 collected by the bottom PMTs cS2$_b$, which has a more homogeneous spatial dependence.
The effect of field distortion is negligible and not implemented in the signal response model because the position correction to account for it is applied to data~\cite{xe1t_sr1}.

Selection criteria were applied to data to ensure good quality of the sample and to optimize the signal-to-background ratio for the dark matter search.
More details of the data selection can be found in~\cite{xe1t_long_analysis_1}.
The detection efficiency loss arises from the software reconstruction efficiency of S1s, and the S1-related and S2-related event selections.
The efficiencies for these are considered as functions of $N_{\mathrm{hit}}$, S1 and S2, respectively, in the signal response model.
In addition, a realistic selection of single scatters is implemented in the simulation.
The rejection of multiple scatters is critical to the search for WIMP signals in XENON1T detector
and is based on the areas of the largest and second largest S2s.
Not all multiple scatters are rejected by this selection.
This is mainly  because of PMT after-pulses, photoionization of impurities in the detector, and the spatial resolution of the detector.
Each energy deposition that is resolvable by the position reconstruction is taken into account in the simulation of the response model.
The same single-scatter selection is applied to the simulated and actual data, in order to accurately address the acceptance of single scatters and the rejection power against multiple scatterings in the response model.

\subsection{Fit to Calibration Data}

The detector response model is constrained using the calibration data from $^{220}$Rn for ER and $^{241}$AmBe and a D-D generator for NR.
Events with cS1 ranging from 0 to 100\,PE are used to constrain the signal response model, covering the cS1 region of interest (3-70\,PE) for WIMP searches~\cite{xe1t_sr0,xe1t_sr1}.
The detector response model obtained from the fit to calibration data 
is used to construct WIMP signal and background models, which are then input to the statistical inference of dark matter search data~\cite{xe1t_sr0,xe1t_sr1}.
The fit is performed simultaneously using all available XENON1T calibration data taken during the first (SR0 with drift field of 120\,V/cm) and second (SR1 with drift field of 81\,V/cm) science data taking periods.
The fit uses the binned likelihood for distributions in log$_{10}$(cS2$_b$/cS1) versus cS1 using the data in the cylindrical fiducial volume defined in SR0~\cite{xe1t_sr0}.
The likelihood is sampled using affine invariant Markov Chain Monte Carlo (MCMC)~\cite{emcee}.
Important nuisance parameters in the detector response model are listed in Table~\ref{tab:response_model_nuisance_parameters}.
There are three parameters for scaling the S1 cut acceptance, S2 cut acceptance, and reconstruction efficiency.
These parameters are constrained by the uncertainties estimated for the three efficiencies, which depend on the signal size.
The $\delta_{\mathrm{s1}}$, $\delta_{\mathrm{s2}}$, $\Delta \delta_{\mathrm{s1}}$, and $\Delta \delta_{\mathrm{s2}}$ in Eq.~\ref{eqn:uncorrected_s1_s2} are signal-size dependent, and thus are not listed in Table~\ref{tab:response_model_nuisance_parameters}.
Their priors can be found in~\cite{xe1t_long_analysis_1}.
We choose to use the more conservative uncertainty between the lower and upper uncertainties given in~\cite{nest_v1.0} for the NEST parameters that describe the response of LXe to NR, except for $\eta$, which parameterizes the Penning quenching of high-energy NRs.
Correlations between the NEST parameters are not provided in~\cite{nest_v1.0} and are, thus, not considered in the priors of this work, in order to be conservative and avoid potential over-constraint on fit parameters.

\begin{figure}[htpb]
    \centering
    \includegraphics[width=\columnwidth]{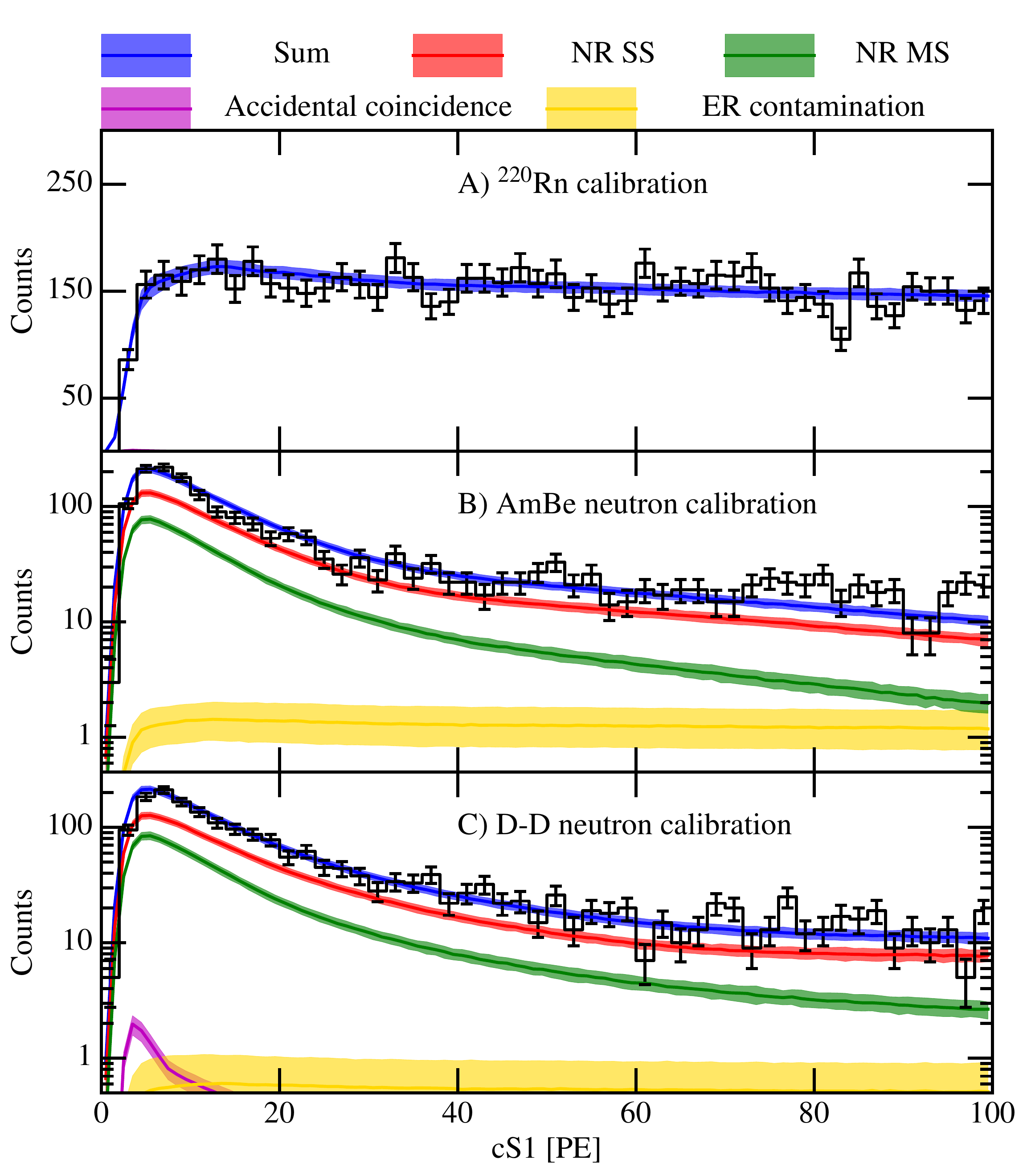}
    \caption{
    The cS1 spectra of the SR1 data (black bars) and the signal response models (blue) for $^{220}$Rn, $^{241}$AmBe, and D-D generator are shown in panel A, B, and C, respectively.
   Solid lines represent the median of the posterior, and the shaded regions show the 15.4\% to 84.6\% credible regions of the posterior.
    The accidental coincidence, ER contamination, single NR scatter (NR SS), and multiple NR scatter (NR MS) components are shown in magenta, gold, red and green, respectively.
    }
    \label{fig:calibration_fit_cs1_spectra_comparison}
\end{figure}

\begin{figure*}[thp]
    \centering
    \includegraphics[width=0.95\textwidth]{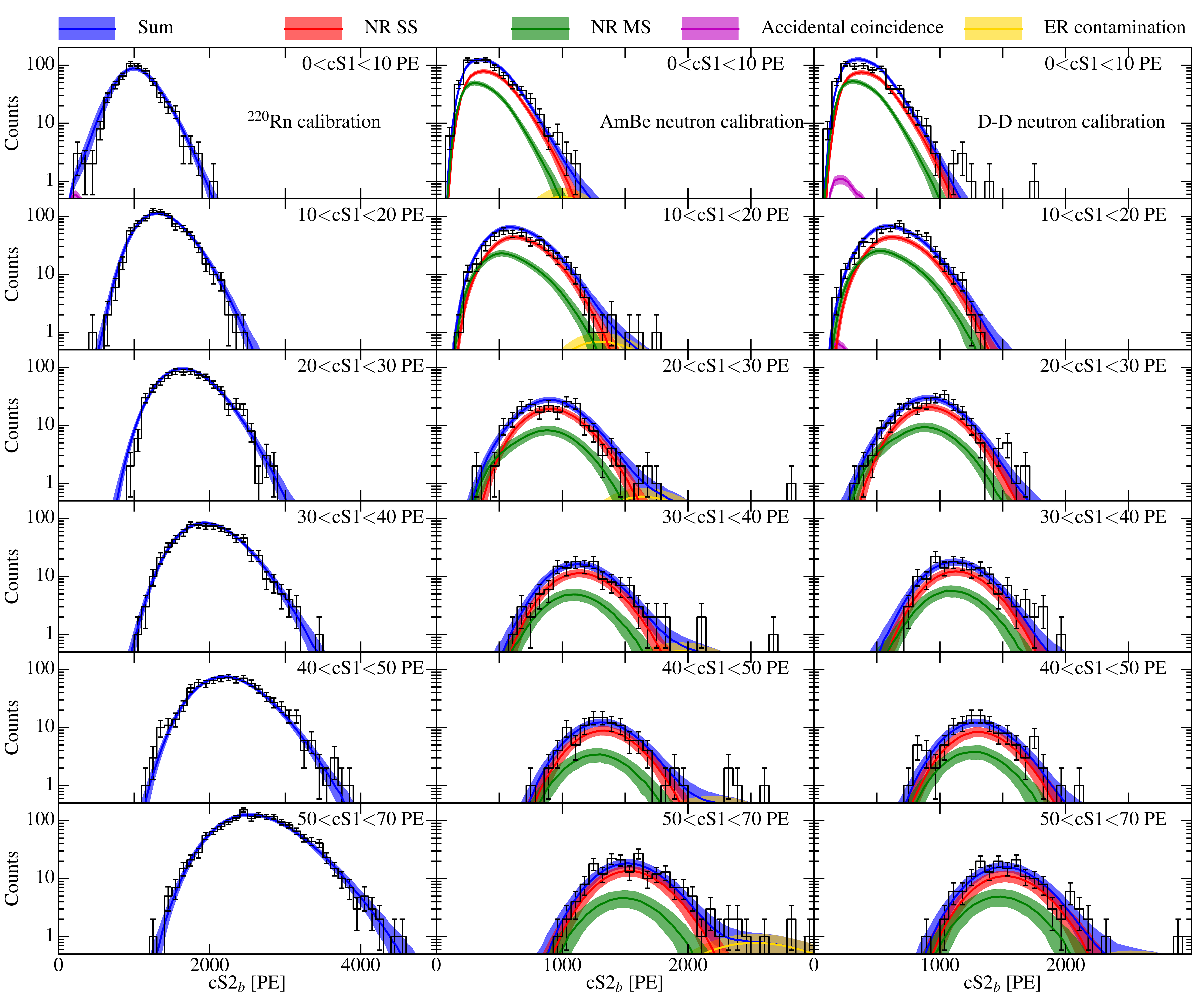}
    \caption{
   The cS2$_b$ distributions in different cS1 ranges of the SR1 data (black bars) and the signal response model (blue) for $^{220}$Rn, $^{241}$AmBe and D-D generator, from left to right, respectively.
   The figure description is the same as in Fig.~\ref{fig:calibration_fit_cs1_spectra_comparison}.
    }
    \label{fig:calibration_fit_cs2_spectra_comparison}
\end{figure*}

Figures ~\ref{fig:calibration_fit_cs1_spectra_comparison} and ~\ref{fig:calibration_fit_cs2_spectra_comparison} show the comparisons of the cS1 spectra and cS2$_b$ spectra, respectively, between the posterior of tested signal response model and data.
For neutron calibrations
the ER contamination are considered in the fit. 
For both ER and NR calibrations, a small fraction of events arise from accidental coincidence (AC) which will be illustrated in section~\ref{sec:background_models}.
The rates of each component are free in the fit, and are not listed in Table~\ref{tab:response_model_nuisance_parameters}.
The matching between the signal response model and the calibration data is good, with the goodness of fit (GoF) p-values (calculated using method in~\cite{goodness_of_fit}) for the cS2$_b$ distributions comparison in different cS1 ranges shown in Table~\ref{tab:gofs_calibration_fit}.
Good agreement between model and data is critical for the WIMP signal and background models, especially for the (cS1, cS2$_b$) region where the WIMP signal is expected.
We also show the p-values for the match of the model with $^{220}$Rn in a reference (cS1, cS2$_b$) region.
The upper cS2$_b$ boundary of the reference region for matching $^{220}$Rn data is defined by the $97.7\%$ percentiles (+2$\sigma$) of NR events in $^{241}$AmBe calibration data.
The GoF  p-values for $^{241}$AmBe and D-D generator data are calculated excluding the lowest S2 region, corresponding to the $0.13\%$ percentile (-3$\sigma$) in S2.
The GoF p-values at different cS1 for matching calibration data are all above the 5\% threshold set for an acceptable fit.


\begin{table}[htp]
    \centering
    \caption{
    Goodness of fit p-values of data-model matching on cS2$_b$ spectra in different cS1 ranges for $^{220}$Rn, $^{241}$AmBe and D-D neutron generator calibration data in SR1.}
    \begin{tabular}{c|cccccc}
    \hline\hline
    \multirow{2}{*}{data} & \multicolumn{6}{c}{cS1 range (PE)} \\
     & 0-10 & 10-20 & 20-30 & 30-40 & 40-50 & 50-70 \\
    \hline
    $^{220}$Rn \textrm{overall} & 0.18 & 0.44 & 0.51 & 0.96 & 0.29 & 0.49 \\
     $^{220}$Rn \textrm{reference} & 0.17 & 0.28 & 0.30 & 0.85 & 0.21 & 0.48 \\
    $^{241}$AmBe & 0.12 & 0.50 & 0.88 & 0.89 & 0.62 & 0.70 \\
    D-D & 0.10 & 0.50 & 0.89 & 0.73 & 0.11 & 0.27 \\
    \hline\hline
    \end{tabular}
    \label{tab:gofs_calibration_fit}
\end{table}

\def\arraystretch{1.3}
\begin{table*}[htpb]
    \centering
    \caption{
    Main nuisance parameters in the detector response model of XENON1T.
    Lower and upper errors of the posterior are calculated as the difference between the median and the 15.4\% and 84.6\% percentiles of the posterior.
    }
    \begin{tabular}{c|cc|cc|c}
    \hline\hline
    \multirow{2}{*}{Par.} & \multicolumn{2}{c|}{Prior} & \multicolumn{2}{c|}{Posterior} &  \multirow{2}{*}{Reference and note} \\
      & SR0 & SR1 & SR0 & SR1 & \\
    \hline\hline
    $W$ & \multicolumn{2}{c|}{13.7$\pm$0.2} & \multicolumn{2}{c|}{13.8$\pm$0.2} & In unit of eV; Eq.~\ref{eqn:fluc_number_of_quanta} \\
    $\langle N_{\mathrm{ex}}/N_i \rangle$ & \multicolumn{2}{c|}{0.06 - 0.20} & \multicolumn{2}{c|}{$0.15^{+0.04}_{-0.06}$} & Eq.~\ref{eqn:fluc_excimer_to_ion_ratio} \\
    $g_1$ & \multicolumn{2}{c|}{0.142$\pm$0.002} & \multicolumn{2}{c|}{0.142$\pm$0.005} &  In unit of PE/photon; Eq.~\ref{eqn:g1_g2_connection} \\
    $g_2$ & \multicolumn{2}{c|}{11.4$\pm$0.2} & \multicolumn{2}{c|}{11.4$\pm$0.2} & In unit of PE/e$^{-}$; Eq.~\ref{eqn:g1_g2_connection}, for cS2${}_\mathrm{b}$ \\
    $p_{\mathrm{dpe}}$ & \multicolumn{2}{c|}{0.18 - 0.24} & \multicolumn{2}{c|}{$0.219^{+0.015}_{-0.023}$} & Eq.~\ref{eqn:g1_g2_connection}, \cite{faham2015measurements,paredes2018response}  \\
    1 - $\tau_\mathrm{e}$/$\tau_\mathrm{e}^{\prime}$ & 0$\pm$0.04 & 0$\pm$0.02 & 0.01$\pm$0.03 & 0.01$\pm$0.01 & Eq.~\ref{eqn:number_of_extracted_electron} and~\ref{eqn:s1_s2_correction} \\
    $\gamma_{\mathrm{er}}$ & \multicolumn{2}{c|}{free} & \multicolumn{2}{c|}{0.124$\pm$0.003} & Eq.~\ref{eqn:er_recomb_parameterization2} \\
    $\omega_{\mathrm{er}}$ & \multicolumn{2}{c|}{free} & \multicolumn{2}{c|}{31$\pm$4} & In unit of keV; Eq.~\ref{eqn:er_recomb_parameterization2} \\
    $\delta_{\mathrm{er}}$ & \multicolumn{2}{c|}{free} & \multicolumn{2}{c|}{0.24$\pm$0.06} & Eq.~\ref{eqn:er_recomb_parameterization2} \\
    $q_0$ & \multicolumn{2}{c|}{free} & \multicolumn{2}{c|}{$1.13^{+0.24}_{-0.32}$} & In unit of keV; Eq.~\ref{eqn:er_recomb_parameterization1} \\
    $q_1$ & \multicolumn{2}{c|}{free} & \multicolumn{2}{c|}{$0.47^{+0.18}_{-0.15}$} & In unit of keV; Eq.~\ref{eqn:er_recomb_parameterization1} \\
    $q_2$ & free & free & 0.041$\pm$0.006 & 0.034$\pm$0.003 & Eq.~\ref{eqn:recomb_fluc_parameterization} \\
    $q_3$ & \multicolumn{2}{c|}{free} & \multicolumn{2}{c|}{1.7$^{+1.3}_{-1.1}$} &  In unit of keV; Eq.~\ref{eqn:recomb_fluc_parameterization} \\
    $\alpha$ & \multicolumn{2}{c|}{1.240$\pm$0.079} & \multicolumn{2}{c|}{1.280$\pm$0.063} & NEST parameters~\cite{nest_v1.0} \\
    $\zeta$ & \multicolumn{2}{c|}{0.047$\pm$0.009} & \multicolumn{2}{c|}{0.045$^{+0.009}_{-0.008}$} & NEST parameters~\cite{nest_v1.0} \\
    $\beta$& \multicolumn{2}{c|}{239$\pm$28} & \multicolumn{2}{c|}{273$^{+24}_{-20}$} & NEST parameters~\cite{nest_v1.0} \\
    $\gamma$ & \multicolumn{2}{c|}{0.0139$\pm$0.007} & \multicolumn{2}{c|}{0.0141$\pm$0.006} & NEST parameters~\cite{nest_v1.0} \\
    $\delta$ & \multicolumn{2}{c|}{0.062$\pm$0.006} & \multicolumn{2}{c|}{0.061$\pm$0.006} & NEST parameters~\cite{nest_v1.0} \\
    $\kappa$ & \multicolumn{2}{c|}{0.139$\pm$0.003} & \multicolumn{2}{c|}{0.138$\pm$0.003} & NEST parameters~\cite{nest_v1.0} \\
    $\eta$ & \multicolumn{2}{c|}{3.3$\pm$0.7} & \multicolumn{2}{c|}{3.3$^{+0.7}_{-0.6}$} & NEST parameters~\cite{nest_v1.0} \\
    $\lambda$ & \multicolumn{2}{c|}{1.14$\pm$0.45} & \multicolumn{2}{c|}{1.15$^{+0.35}_{-0.27}$} & NEST parameters~\cite{nest_v1.0} \\
    $\epsilon_{\mathrm{ext}}$ & 96\% & 96\% & - & - & Fixed; Eq.~\ref{eqn:number_of_extracted_electron} \\
    $\Delta G/G$ & 0.24 & 0.25 & - & - & Fixed; Eq.~\ref{eqn:number_of_proportional_light}\\
    $\nu_{d}$ & 0.144 & 0.134 & - & - & In unit of cm/$\micro$s; fixed; Eq.~\ref{eqn:number_of_extracted_electron}\\
    $F$ & 120 & 81 & - & - & In unit of V/cm; \\
    \hline\hline
    \end{tabular}
    \label{tab:response_model_nuisance_parameters}
\end{table*}
\def\arraystretch{1.0}%

\section{Background and Dark Matter Signal Models}
\label{sec:background_models}

Background and DM signal models are crucial in the statistical interpretation of the DM search results in XENON1T.
There are four background components in XENON1T: ER, NR, surface, and accidental coincidence (AC). 
The ER and NR background models, as well as the WIMP signal model, are constructed based on the detector response model illustrated in section~\ref{sec:detector_signal_response_model} (we call the detector response model ``posterior'' after fitting to the calibration data).
The surface and AC background models are constructed using data-driven methods.
The background and WIMP signal models are 3-D distributions in cS1, cS2$_b$, and the spatial coordinate of the detector.
The background and DM signal models also include the uncertainties in cS1, cS2$_b$ and spatial distribution, as well as in the absolute rate of the background and DM signal.
In this section, the details of the background and DM signal models used in the statistical inference of XENON1T results~\cite{xe1t_sr0, xe1t_sr1} are given.

The final dark matter search is performed between 3$<$cS1$<$70\,PE, and 50.1$<$cS2$_b$$<$7940\,PE. Below 3\,PE, the S1 acceptance is very small due to the 3-fold PMT coincidence requirement for S1s. The upper \cSone cut is chosen to contain most of spin-independent WIMP recoil spectra, shown in Fig.~\ref{fig:wimp_signal_model}.
In previous XENON analyses, fiducial volumes in radius R and $z$ were constructed to  provide a low background for the analysis. With the inclusion of a model for the surface background, presented in Section~\ref{sec:surface}, the analysis could be extended to also consider the radius as an analysis variable. The analysis volume is defined by a maximal radius, $42.8~$cm, and a R-dependent $z$-cut, shown in Fig.~\ref{fig:ms_neutron} with a magenta line. The construction of this cut, which was made to include regions of the detector where the total background rate was approximately constant with $z$, is presented in~\cite{xe1t_long_analysis_1}.
%

The magnitude of the radiogenic background, discussed in Section~\ref{sec:nrbkg}, is attenuated moving towards the center of the detector. In order to optimize the discovery power of the analysis, a partition of the detector, with a clean, "core" volume was proposed.
Optimizing for discovery significance yielded a central $0.65$ tonne volume, shown with a dashed green line in Fig.~\ref{fig:ms_neutron}. 
The expected radiogenic neutron rate in this volume is $36\%$ of the average rate in the analysis volume. 

\subsection{Electronic Recoil Background Model}
\label{subsec:ER}

Although XENON1T achieved an excellent discrimination power between the ER background and NR signal, with an average ER leakage fraction below the NR median of about 0.3\%~\cite{xe1t_sr1}, the ER component is the dominant background for the DM search due to its high rate in comparison with the other background sources.

In the energy region of interest for WIMP search ($<$100\,keV$_\mathrm{NR}$), the dominant component contributing to the ER background are $\beta$-decays of $^{214}$Pb. The 
$^{214}$Pb is a progeny of $^{222}$Rn, which is emanated from $^{238}$U daughters in the detector materials, and can convect and diffuse into the inner volume of the detector.
Decays of $^{218}$Po and $^{214}$Bi-$^{214}$Po, which are also $^{222}$Rn progenies, can be identified to estimate the rate of $^{214}$Pb.
This selection is based on the unique energy and time profiles of $^{218}$Po and $^{214}$Bi-$^{214}$Po decays, and gives activities of 71$\pm$5(stat.)$\pm$7(sys.) and 29$\pm$3(stat.)$\pm$3(sys.) events/ton/year/keV (tyu), respectively.
The difference between the activities of $^{222}$Rn progenies is likely due to their plate-out onto the electrode and polytetrafluoroethylene (PTFE) reflector surfaces.
As the $^{214}$Pb decay occurs between the $^{218}$Po and $^{214}$Bi decays, the rate of $^{214}$Pb is in range of 29 to 71\,tyu which is consistent with the estimate of 56$\pm$6(stat.)$\pm$6(sys.)\,tyu from~\cite{xe1t_mc}, where 10\,$\micro$Bq/kg of $^{222}$Rn were assumed.

The second largest component of the ER background is $\beta$-decays of $^{85}$Kr.
The concentration of natural krypton $^\mathrm{nat}$Kr/Xe was reduced to 0.36$\pm$0.06\,ppt by the end of SR0 through cryogenic distillation~\cite{xe1t_distillation}.
With regular rare-gas mass spectrometry measurements~\cite{xe1t_rgms} during SR1, we measured an average natural krypton concentration $^\mathrm{nat}$Kr/Xe of 0.66$\pm$0.11\,ppt, resulting in an average decay rate for $^{85}$Kr of 7.7$\pm$1.3\,tyu,  using the conversion derived from data with high concentration of krypton at the beginning of XENON1T operation.
These high-krypton concentration data also gave a $^{85}$Kr/$^\mathrm{nat}$Kr ratio of (1.7$\pm$0.3)$\times$10$^{-11}$ mol/mol.
Taking the ER contributions from material radioactivity, solar neutrino scatterings, and $\beta$-decays of $^{136}$Xe of 8$\pm$1, 2.5$\pm$0.1, and 0.8$\pm$0.1\,tyu, respectively, into account, the total ER background rate in the region of interest (ROI) for DM searches is estimated to be between 48$\pm$5 and 90$\pm$8\,tyu.
This is consistent with the prediction of 75$\pm$6 (sys.)\,tyu from~\cite{xe1t_mc} and with the best-fit of 82$_{-3}^{+5}$(sys)$\pm$3(stat)\,tyu of low energy ER background from~\cite{xe1t_sr1}.

\begin{figure}[htp]
    \centering
    \includegraphics[width=0.95\columnwidth]{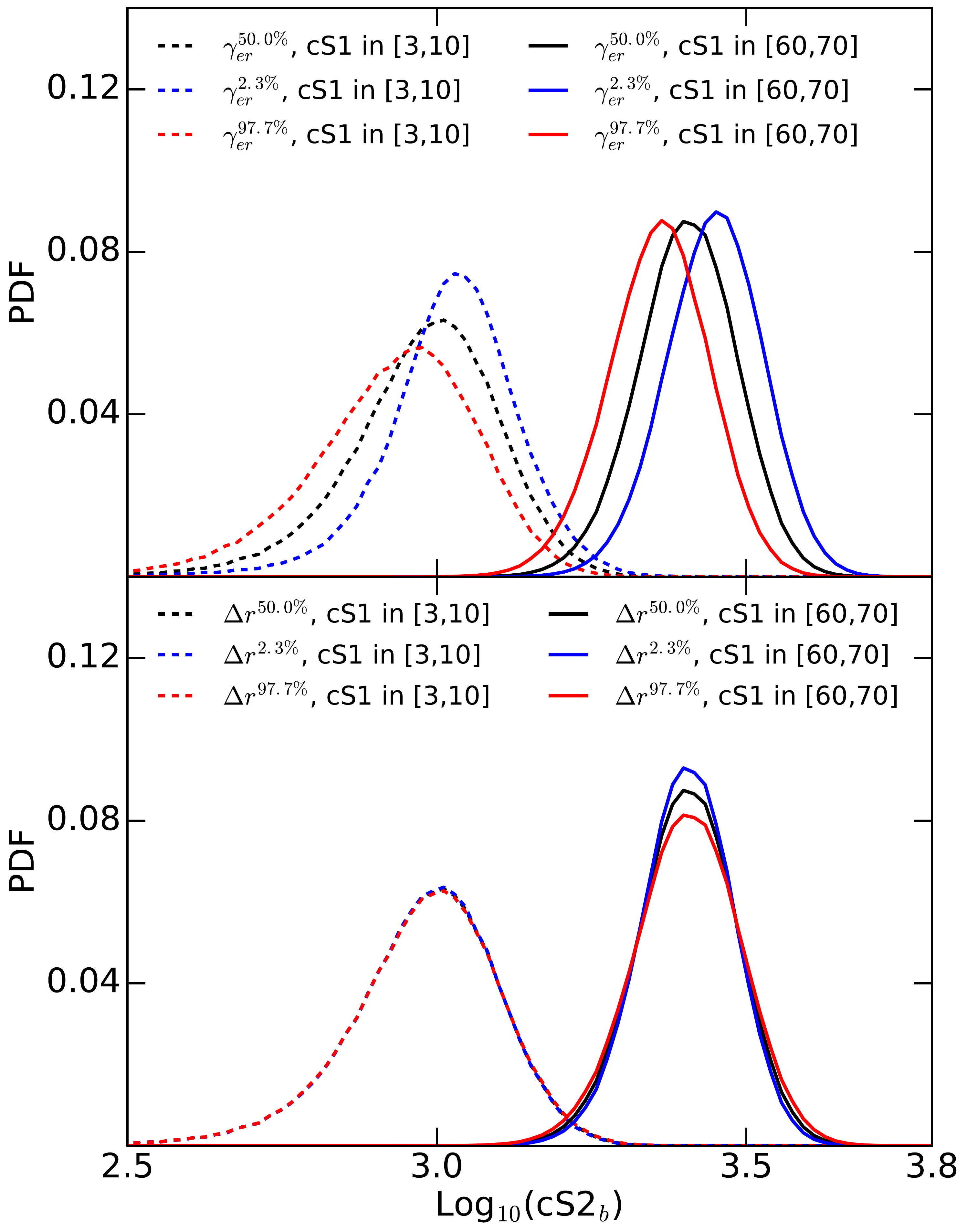}
    \caption{
    Variations in log$_{10}$(cS2$_b$) distributions as $\gamma_{er}$ (top panel) and $\Delta r$ (bottom panel) vary, from the 2.3\% to 50.0\%, and then to 97.7\% percentile of the signal model posterior, in different cS1 ranges.
    }
    \label{fig:er_template}
\end{figure}

The energy distribution of the ER background in the ROI is assumed to be uniform due to the dominance of the flat spectrum from $^{214}$Pb beta decay.
Uncertainties in the (cS1, cS2$_b$) distribution for the ER background are dominated by the uncertainties in $\langle r \rangle$, and its fluctuations, $\Delta r$. 
The uncertainty in $\langle r \rangle$ is mainly from parameter $\gamma_{\mathrm{er}}$ in Eq.~\ref{eqn:er_recomb_parameterization2}.
The effects of varying $\gamma_{\mathrm{er}}$ and $\Delta r$ are shifting the mean and changing the spread, respectively, of the ER distribution in cS2$_b$.
Fig.~\ref{fig:er_template} shows the variation of the ER distributions on log$_{10}$(cS2$_b$) in different cS1 ranges.
The distributions are produced based on the detector response model with the rest of the nuisance parameters (shown in Table~\ref{tab:response_model_nuisance_parameters}) marginalized to the point estimation (median posterior).
Due to the computational complexity, the variation of background model in terms of (cS1, cS2$_b$) distribution in the statistical interpretation is practically interpolated using the distributions that are computed at 2.3\%, 6.7\%, 15.9\%, 30.9\%, 50.0\%, 69.1\%, 84.1\%, 93.3\%, 97.7\% percentiles of the posterior, which correspond to -2, -1.5, -1, -0.5, 0, 0.5, 1.0, 1.5, 2.0 $\sigma$, respectively, in standard deviations.
Given that the ER background induced by radioactivities in detector materials is subdominant, the ER background is assumed to be spatially uniform inside the analysis volume.

\subsection{Nuclear Recoil Background Model}
\label{sec:nrbkg}
The NR background, which has a similar (cS1, cS2$_b$) distribution to the WIMP signal, contributes 1.43 events to the 1\,tonne$\times$year exposure data~\cite{xe1t_sr1}.
Radiogenic neutrons, muon-induced neutrons and solar neutrinos contribute to this NR background.


Radiogenic neutrons are generated by ($\alpha$, n) reactions and spontaneous fissions of material radioactive impurities, and are the largest source of NR background.
The neutron yields of materials are predicted using SOURCES-4A~\cite{sources_4a} based on the measured radioactivity of the detector materials~\cite{aprile2017xenon1t}.
The propagation of generated neutrons is simulated using the GEANT4 toolkit~\cite{agostinelli2003s}.

\begin{figure*}[bhtp]
    \centering
    \includegraphics[width=0.95\columnwidth]{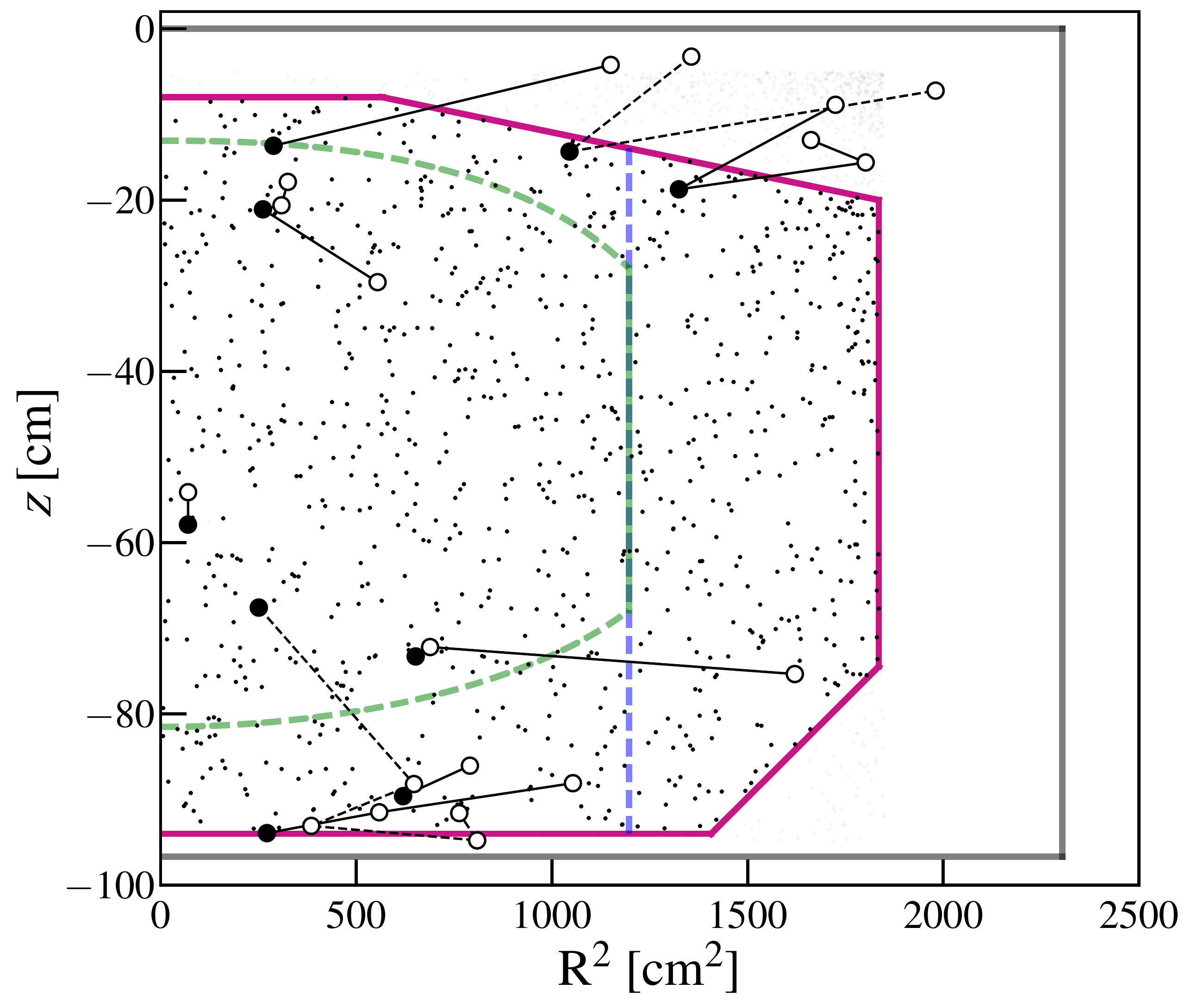}
    \includegraphics[width=0.95\columnwidth]{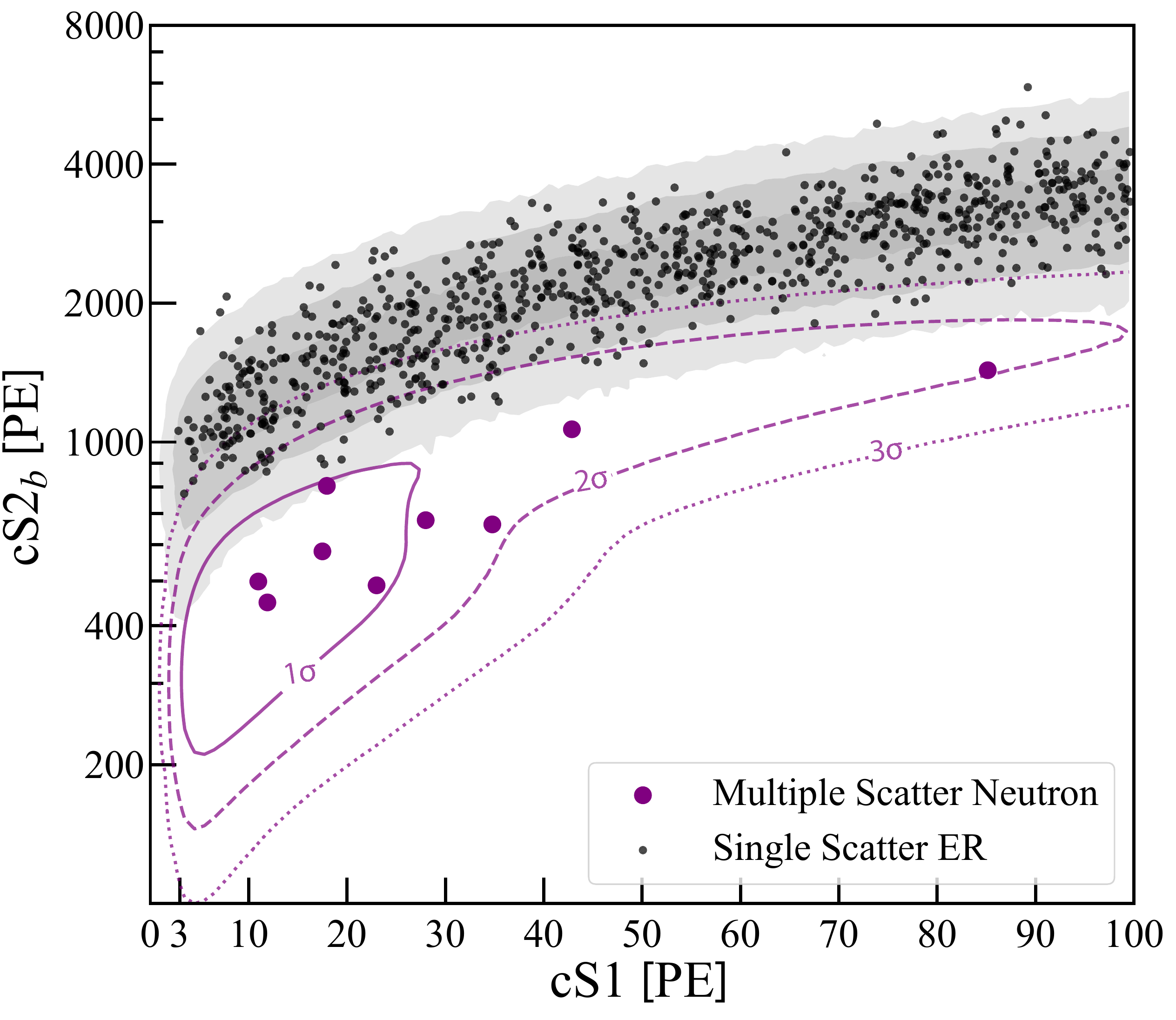}
    \caption{
    Spatial (left panel) and (cS1, cS2$_b$) (right panel) distribution of identified multiple neutron scatters in SR0 and SR1 DM search data, along with blinded DM search data (ER events) shown in black dots.
    The solid circles in the left panel represent the primary (largest cS2$_b$) scatter positions while the hollow circles show the positions of the secondary scatters.
    The solid or dashed black line connects the scatters that are in the same event.
    XENON1T TPC boundaries are shown as solid gray lines.
    Different volumes defined in~\cite{xe1t_sr1}, 0.65\,t (dashed green), 0.9\,t (dashed blue) and 1.3\,t (solid magenta), are shown.
    In the right panel, the 1$\sigma$, 2$\sigma$, and 3$\sigma$ contours of the expected distribution of neutron multiple scatters are shown as solid, dashed and dotted purple lines, respectively.
    Solid purple circles show the primary (cS1, cS2$_b$) of the identified neutron multiple scatters in DM data.
    As a comparison, the shaded black regions display the 1$\sigma$ (dark) and 2$\sigma$ (light) probability density percentiles of the ER background component for SR1.
    }
    \label{fig:ms_neutron}
\end{figure*}

Because of the large uncertainty ($\sim$50\%) in the estimated neutron rate, multiple neutron scatter events in DM search data and calibration data are used to further constrain the rate uncertainty. 
For this purpose, multiple scatter events are unblinded prior to single scatters.
Nine neutron multiple scatter events were identified in the DM search data (SR0 and SR1 combined) within the 1.3-tonne fiducial volume (FV).
After a study comparing the single-to-multiple scatter ratio in data and simulations, the data-constrained neutron rate is estimated to be 55$_{-20}^{+33}$ neutron scatters, including single and multiple scatters, per year in the 2-tonne active volume of XENON1T detector (the expectation is 37 neutron scatters per year from simulation).
About 35\% of the selected single scatters by radiogenic neutrons are misidentified multiple scatters.
Fig.~\ref{fig:ms_neutron} shows the spatial and (cS1, cS2$_b$) distributions of the identified neutron multiple scatters in DM search data together with the single scatter events surviving from the blinding cut~\cite{xe1t_sr1}.

Neutrons emitted from bottom PMTs in the bottom array have a probability to  scatter in the region between the TPC cathode and PMTs, referred to as the "below-cathode" region.
The scintillation light from these scatters is detected, but electrons are lost because the electric field in the below-cathode region drifts them away from the active volume.
Such events, with at least one scatter in the below-cathode region and a single scatter in the 1.3\,tonne FV, are named neutron-X events, and have a lower cS2$_b$ to cS1 ratio than standard neutron scatters.
The relative ratio and difference in (cS1, cS2$_b$) distributions, as well as in spatial distributions, between normal neutron scatters and neutron-X events is taken into account in the simulation for building the NR background model.

Muon-induced neutrons are estimated to be subdominant to radiogenic neutrons in the active volume~\cite{xe1t_mc}, and the rate of muon-induced neutrons is further suppressed by applying a muon-veto selection. The final rate is approximately 2 orders of magnitude smaller than that of the radiogenic neutrons, and its contribution is neglected in the NR background modelling.


Compared with~\cite{xe1t_sr0}, the NR background model induced by solar neutrino scatters has been updated with the latest results from COHERENT~\cite{akimov2017observation}, and is 23\% smaller.
On top of the neutrino flux uncertainty of 14\%, the updated NR background model also takes into account the uncertainty of about 21\% from~\cite{akimov2017observation} and of about 22\% from the signal response model at the energy region of interest for neutrino scatters.
The resulting rate uncertainty for solar neutrino induced coherent elastic neutrino-nucleus scatters (CE$\nu$NS) is 34\%.

\subsection{Surface Background Model}
\label{sec:surface}
Interactions in the bulk region of the detector can be modelled combining knowledge of LXe and detector-related responses, as discussed in the previous section for the ER and NR background models. 
For background events where the knowledge of the response is incomplete or missing, data-driven methods have to be developed to model the events.
In this and the following section, two classes of such background components are analyzed for the XENON1T WIMP searches, events originating from the detector surface, and accidental coincidences of unrelated S1 and S2 signals.

Several experiments~\cite{cdms_surface,deap} have demonstrated that detector surfaces exposed to ambient air during construction are contaminated by a large amount of radon progeny, in particular $^{210}$Pb. With a 22\,y half-life, $^{210}$Pb decays at a constant rate within the lifetime of the XENON1T experiment. For WIMP searches, ion recoils of $^{206}$Pb from $^{210}$Po $\alpha$-decays, $\beta$-decays and the resulting X-rays and Auger electrons of $^{210}$Pb are particularly important. Due to incomplete knowledge of LXe responses and detector physics in presence of PTFE, as well as complicated decay structures, a full model including the relevant physics processes has not yet been achieved in XENON1T. Instead, a data driven approach is adopted to predict the distribution of this background. 

\begin{figure}[tbp]
\centering
\includegraphics[width=\columnwidth]{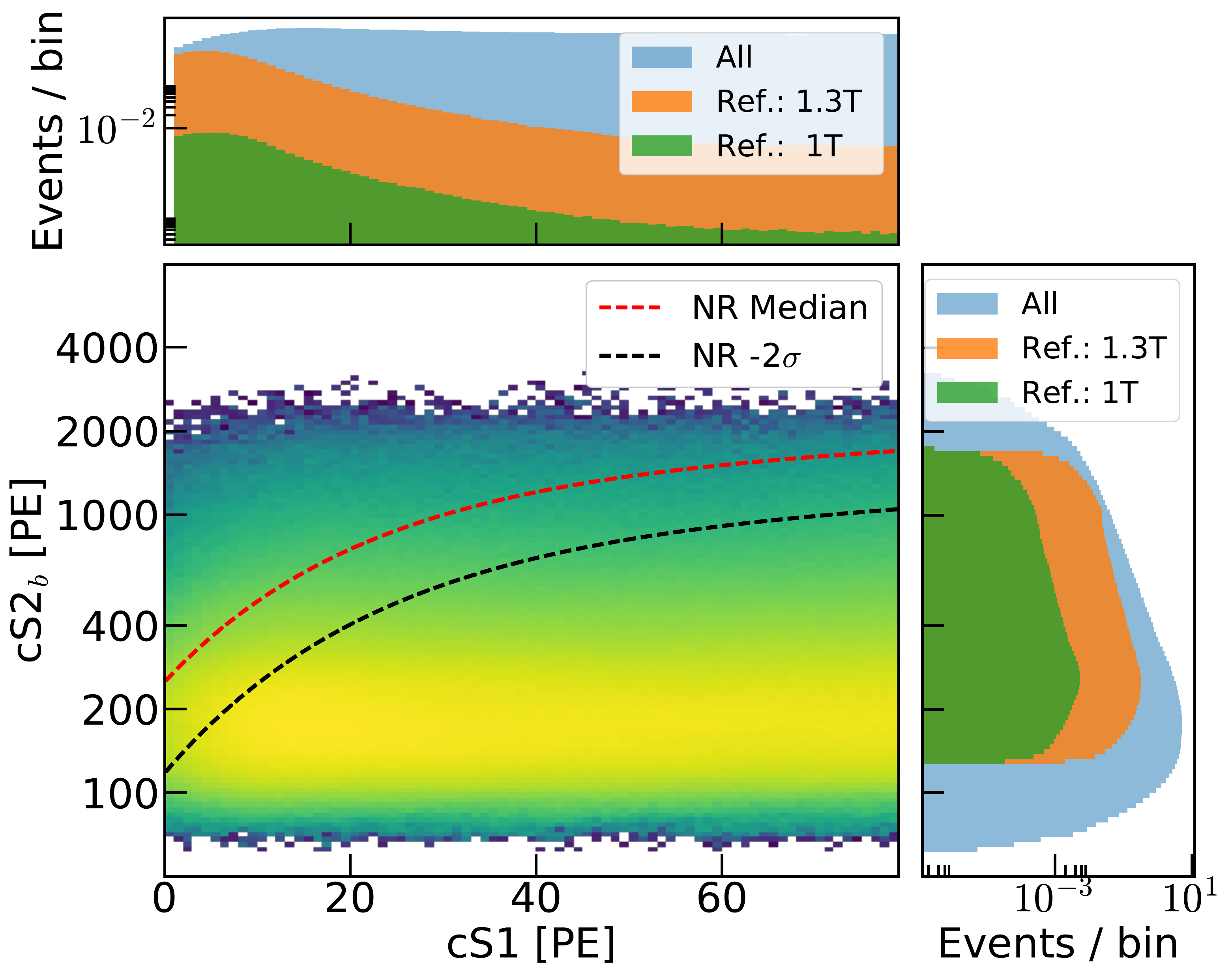}
\caption{\label{fig:surface_bg} 
Illustration of the surface background distributions in cS1 and Log$_{10}$(cS2$_b$), with projections on each axis showing the expected distribution within the entire analysis space (blue), and in the reference region for the entire 1.3\,tonne FV (brown), and the inner 1 tonne (green). The reference region lies between the NR median and $-2\sigma$ quantile lines, marked by red and black lines, respectively. 
}
\end{figure}

Background from radon progeny was modelled in cS1, cS2$_b$, S2, R, and $z$ spaces, with a distribution $f_{\rm{Surf}}(\mathrm{cS1, S2, cS2}_b, \mathrm{R}, z)$. Radial position of surface events are reconstructed nearly symmetrically around the TPC boundary, with an uncertainty determined by the S2. Events mis-reconstructed outside the TPC are used to model the background distribution in S2, cS2$_b$, cS1, and $z$, denoted as $f_{\mathrm{Surf-1}}(\mathrm{cS2}_b, \mathrm{S2, cS1}, z)$, with a kernel-density-estimation method~\cite{scikit-learn}. The distribution in cS2$_b$ and cS1 is shown in Fig.~\ref{fig:surface_bg}. Due to significant charge losses at the PTFE panels, the surface background overlaps significantly with the nuclear recoil region of interest (between nuclear recoil median and $-2\sigma$ quantile). In contrast, the R distribution provides excellent rejection power. 
To construct the distribution of surface background in R and S2 space, events originating at the PTFE surface are selected as control sample with an S1 size out of the region of interest, including the $^{210}$Po $\alpha$-decays. In each S2 slice, the radial distribution of the control-sample events is fitted, including an uncertainty estimated by different fitting functions. The 2-dimensional distribution of surface background, denoted as $f_{\mathrm{Surf-2}}(\mathrm{R, S2})$, is combined with $f_{\mathrm{Surf-1}}(\mathrm{cS2}_b, \mathrm{S2, cS1}, z)$ to form a complete model of the surface background as $f_{\mathrm{Surf}}(\mathrm{cS1, S2, cS2}_b, \mathrm{R}, z)$, including the uncertainty in the radial distribution.

The total rate is normalized to the number of events reconstructed outside the TPC boundary. 
In the WIMP search~\cite{xe1t_sr1}, a radius cut of 42.8\,cm is placed to reduce the surface background to $\sim100$ events. In the likelihood fit described in section~\ref{sec:statistical_inference}, the total surface background expectation value is conservatively treated as a free parameter. 

\subsection{Accidental Coincidence Background Model}

The accidental coincidence (AC) of uncorrelated S1s and S2s (referred to as lone-S1 and lone-S2, respectively) is the fourth background component considered in the XENON1T model.
Lone-S1s and -S2s originate from energy depositions in the non-active regions of the detector, where the scintillation or ionization signal are not detectable.
For example, an energy deposition in the below-cathode region does not produce an ionization signal, and a deposited energy very close to the gate may have its scintillation signal blocked by the mesh. Similar to the surface background model described above, AC background is constructed through a data-driven approach by random pile-up of lone-S1 and lone-S2 samples from data.

The lone-S1 sample is obtained by searching for S1s in the time window before the larger primary S1 in each digitized event waveform. The estimated lone-S1 rate ranges from 0.7 to 1.1~Hz, depending on the requirement of noise rate in the search window. The difference in lone-S1 rate between science runs is negligible.
The lone-S2 sample is obtained using events with no S1 found in the digitized waveform, or with a reconstructed $z$ position larger than the maximum drift time. The observed lone-S2 rate (with S2 threshold of 100\,PE) is determined as 2.6~mHz and stays constant during both science runs. Selection criteria are applied in the lone-S1 and lone-S2 samples directly, while the event selections involving S2 (S1) are excluded. 
The AC event rate is calculated as $\mathrm{R_{AC} = R_{lS1} \cdot R_{lS2}} \cdot \Delta t$, where $\mathrm{R_{lS1}}$ and $\mathrm{R_{lS2}}$ are the rates of lone-S1s and lone-S2s, respectively.
$\Delta t$ is the coincidence time window, which is fixed to the maximum drift time in the TPC (674~$\micro$s for SR0 and 727$~\micro$s for SR1). 
The AC background is nearly uniformly distributed in ($x$, $y$, $z$) position space. 
This yields a total AC background of 44.2 events in SR1, which is reduced to 7.0 events with an S2 threshold of 200\,PE. Several selection criteria were developed to suppress AC background to a sub-dominant contribution and their rejection efficiency were estimated by simulation. 

\begin{figure}[tbp]
\includegraphics[width=\columnwidth]{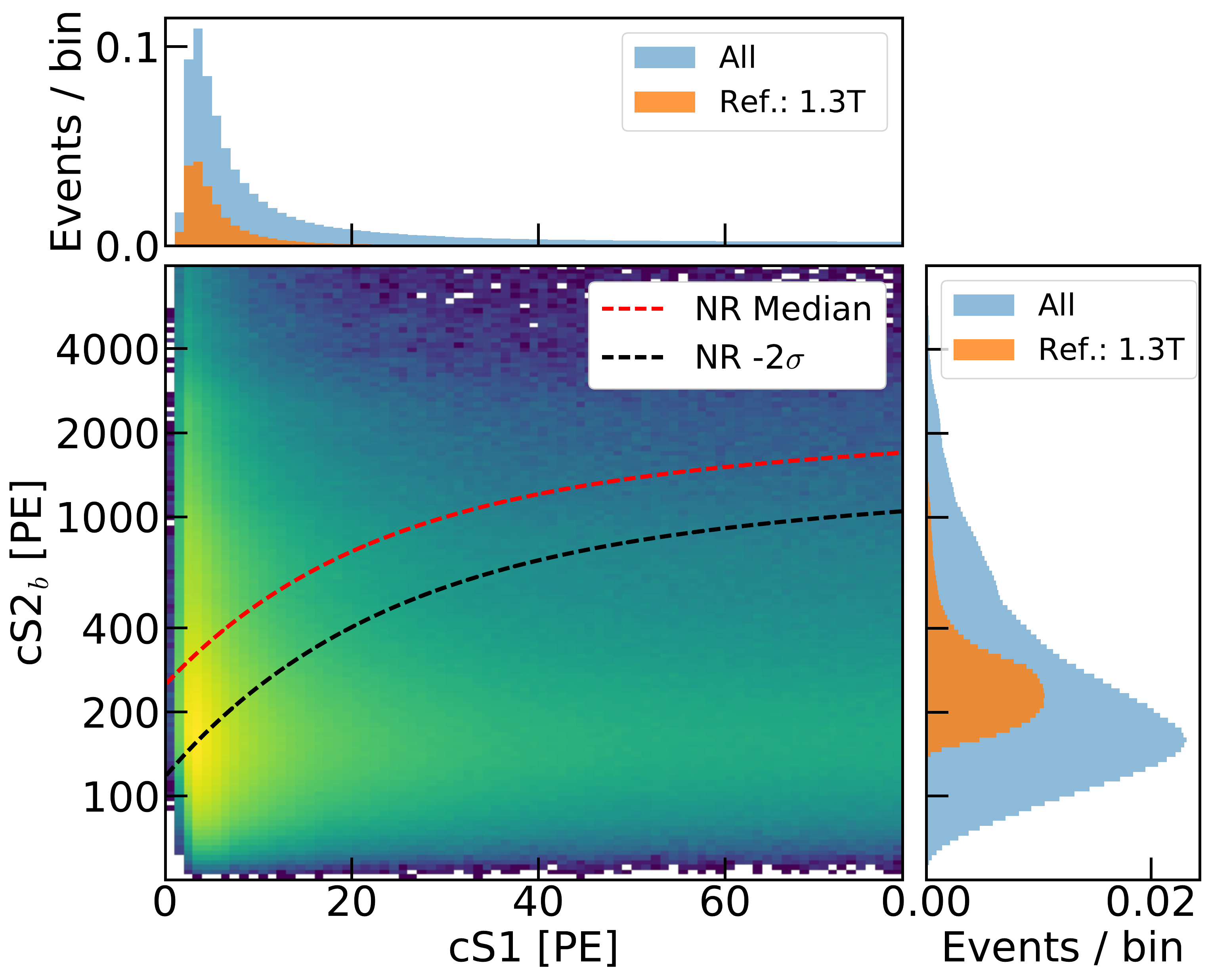}
\caption{\label{fig:ac_bg} 
Illustration of the accidental coincidence background distribution in cS1 and log$_{10}$(cS2$_b$), with projections on each axis showing the expected distribution within the entire analysis space (blue), and in the reference region for 1.3\,tonne FV. The reference region lies between the NR median and $-2\sigma$ quantile lines, marked by red and black lines, respectively. 
}
\end{figure}

We simulate the distribution of AC events by sampling and randomly pairing the lone-S1 and -S2 samples. Interaction positions of the event are calculated by applying a field distortion correction to the sampled position. The correction of S1 and S2, depending on the event position, as well as event selections are all applied to the simulated sample. Examples are the drift-time dependent S2 width and S1 fraction in top PMT array cuts. The final AC background prediction is done with the simulation sample where all selection criteria are applied. The AC model was validated before unblinding, using both $^{220}$Rn calibration data and WIMP search data outside the ROI (for example the sample with S2 between 100 and 200\,PE, or with S1s being identified as single electron S2s). The predicted AC rate after all selection criteria in the ROI for SR0 and SR1 combined is 0.47\,-\,0.74 events per tonne per year.
As shown in Fig.~\ref{fig:ac_bg}, the AC distribution is concentrated in the low cS1 region, making it important in the search for light WIMPs.

\subsection{WIMP Signal Model}

\begin{figure}
    \centering
    \includegraphics[width=0.98\columnwidth]{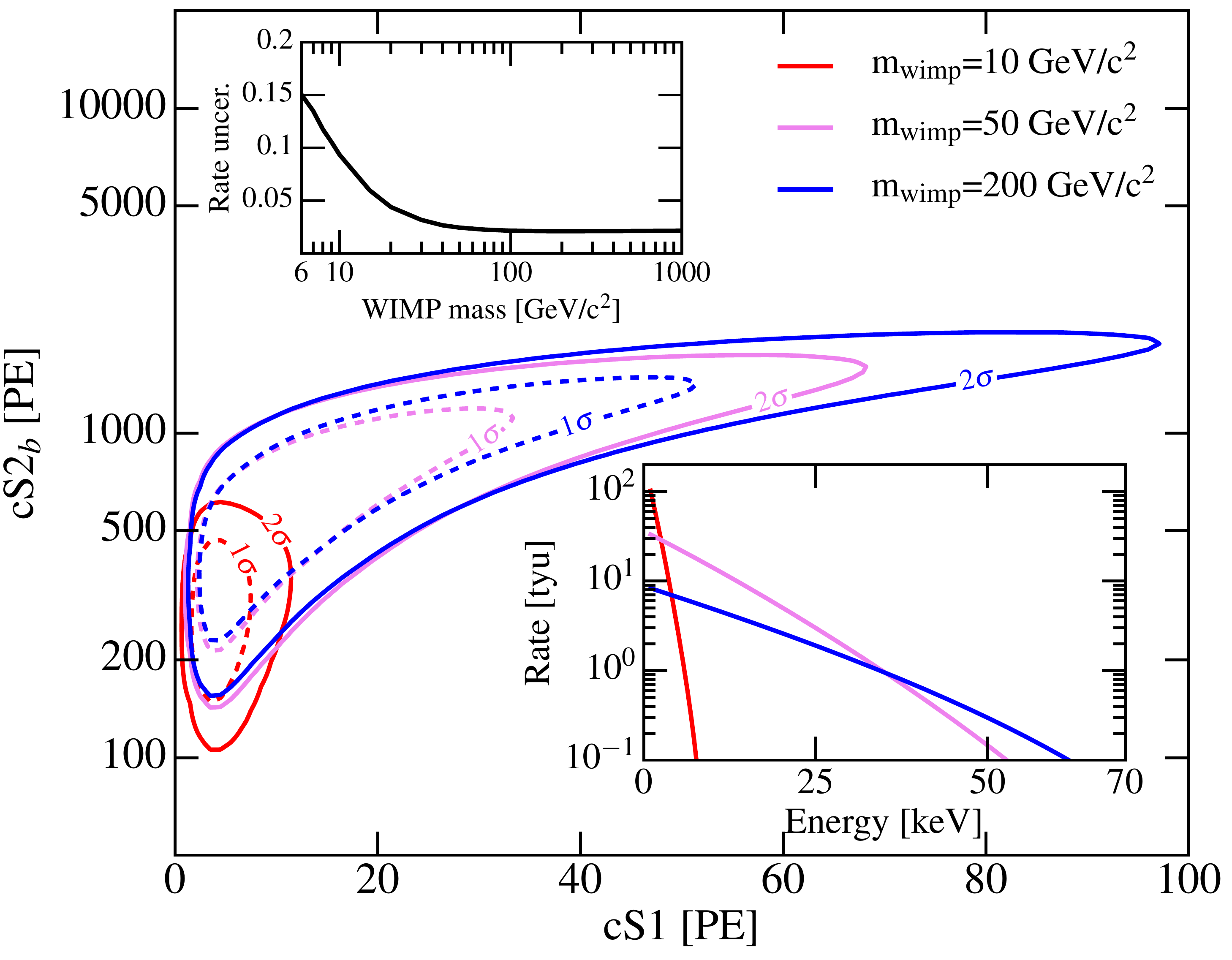}
    \caption{
    The main panel shows the 1$\sigma$ (dashed line) and 2$\sigma$ contours of the WIMP signal model in (cS1, cS2$_b$) space, for WIMP mass of 10\,GeV/c$^2$ (red), 50\,GeV/c$^2$ (violet), and 200\,GeV/c$^2$ (blue), respectively.
    The lower-right inset shows the differential energy rate, which is in unit of tyu, for these three WIMP masses, with an assumed WIMP-nucleon cross section of 10$^{-45}$cm$^2$.
    The upper-left inset shows the rate uncertainty of WIMP signal model as a function of WIMP mass.
    }
    \label{fig:wimp_signal_model}
\end{figure}

The WIMP signal model depends on the dark matter mass, and assumes a uniform distribution of WIMP signals in the FV.
The signal energy spectrum is computed in the same way as in~\cite{xe1t_sr0,xe1t_sr1}.
Distributions in (\cSone, \cStwob), as well as the energy spectra for 10, 50, and 200\,GeV/c$^2$ WIMPs are shown in Fig.~\ref{fig:wimp_signal_model}.
The uncertainty in the WIMP signal model in the (cS1, cS2$_b$) distribution, which comes from the uncertainty in the detector response model, is sub-dominant to the uncertainties of background models in the statistical inference. Therefore, we approximate the uncertainty of the WIMP signal model only in the form of its rate uncertainty.
The approximated rate uncertainties as a function of WIMP mass are shown in Fig.~\ref{fig:wimp_signal_model}.

\section{Inference of XENON1T Data}
\label{sec:statistical_inference}
\newcommand{\lagr}{\mathcal{L}} 
\newcommand{\LL}{\log \mathcal{L}} 
\newcommand{\sr}{\mathrm{SR}}
\newcommand{\nuis}{\theta}
\newcommand{\nuiss}{\boldsymbol{\nuis}}
\newcommand{\nuissh}{\hat{\boldsymbol{\nuis}}}
\newcommand{\nuisshah}{\hat{\nuissh}}
\newcommand{\Poiss}[2]{\mathrm{Pois}(#1 | #2)}
\newcommand{\Gauss}[3]{\mathrm{Gaus}(#1 | #2,#3 )}
\newcommand{\Uniform}[3]{\mathrm{Uniform}(#1 | #2,#3)}
\newcommand{\xmeas}{\vec{x}}
\newcommand{\pdf}{\textsc{pdf}}

\newcommand{\ud}{\mathrm{d}}

\newcommand{\rnttz}{${}^{220}$Rn}
\newcommand{\cevns}{CE$\nu$NS}
\newcommand{\srz}{SR0}
\newcommand{\sro}{SR1}


\newcommand{\amm}{a_\mathrm{mm}}
\newcommand{\bmm}{b}


In this section we describe the general techniques used in XENON1T for hypothesis testing and construction of confidence intervals. 
The focus is on the description of the specific likelihood used for the statistical interpretation of the XENON1T 1~tonne-year WIMPs search data~\cite{xe1t_sr1,xenon_1t_sd}.

\subsection{Hypothesis Testing and Confidence Intervals}
\label{inferencemethod}
The profile log-likelihood ratio is used as the test statistic for both confidence intervals and discovery assessment,
\begin{equation}
   q(\sigma) \equiv -2\cdot\log\frac{\lagr(\sigma,\nuisshah)}{\lagr(\hat{\sigma},\nuissh)},
   \label{eqn:stat1}
\end{equation}
where $\lagr$ stands for the XENON1T likelihood defined in Section~\ref{subsec:likelihood}, 
$\hat{\sigma}$ and $\nuissh$ are the signal and nuisance parameters that maximize the likelihood 
overall, and $\nuisshah$ are the nuisance parameters that maximize the likelihood with the 
condition that the signal strength is $\sigma$. To avoid unphysical regions, the best-fit signal 
$\hat{\sigma}$ is constrained to be non-negative. 

A signal hypothesis $H_{\sigma}$ is tested against the data by computing the p-value of the observed test statistic $q(\sigma)$ given $H_{\sigma}$.
In case of relatively large expected signal-like background, asymptotic formulae~\cite{CowanAsymptotic,Wilks} for the distribution of $q(\sigma)$ are convenient. 
However, given the low background of XENON1T, it was found that these approximations no longer hold. This led to an under-coverage of cross-section and resulted in a overestimated limit of $10\%$ on average in the SR0 results~\cite{xe1t_sr0}. 
Therefore, the distribution of $q(\sigma)$ is computed using toy Monte Carlo (toy-MC) simulations of the background and signal models. Note that in computing these distributions the auxiliary measurements related to each nuisance parameter are also varied per toy-MC data sample.

\begin{figure}[htbp]
    \centering
    \includegraphics[width=\columnwidth]{{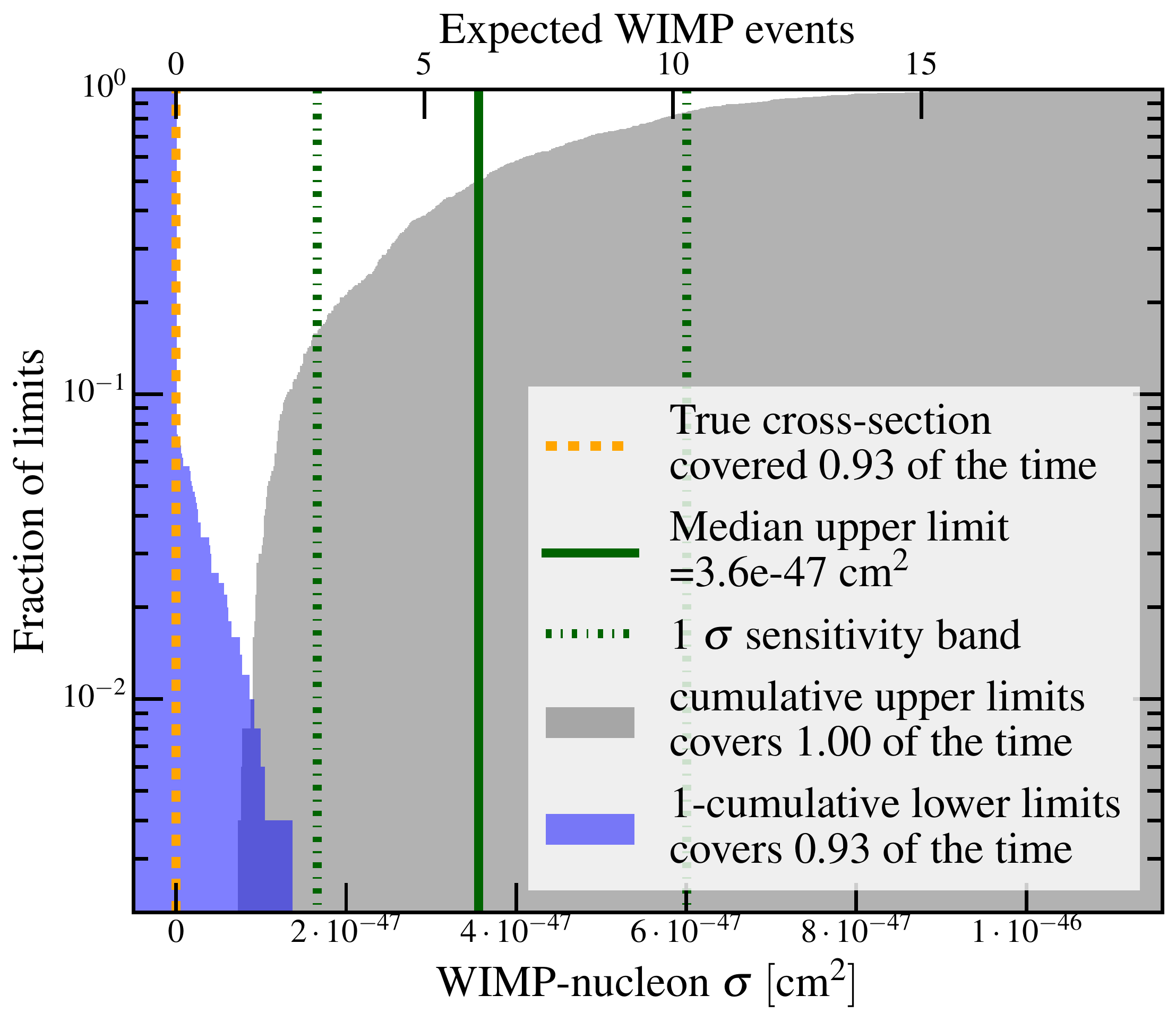}}
    \caption{Distribution of the lower (blue) and upper (gray) toy-MC limits on a 50~\gevcsq~WIMP cross-section. The upper (lower) limits are drawn as a cumulative (complementary cumulative) distribution, to show the fraction of limits that include a certain cross-section. The corresponding signal expectation value is shown on the upper horizontal axis. The toy-MC did not include a true signal, indicated with a dashed orange line at $0$.  All upper limits are above $0$, and $0.93$ of lower limits are equal to $0$, giving a total estimated coverage of $0.93$ for the 500 toy-MCs used in this example, part of the data-set used for figure~\ref{fig:coverage}.
    As the cross-section is constrained to be non-negative, the survival fraction below 0 is $1$.
    Green solid and dot-dashed lines show the median upper limit and $1~\sigma$ sensitivity band. 
    }
    \label{fig:sensi_construction}
\end{figure}

A Feldman-Cousins construction in the profile likelihood~\cite{FeldmanCousins,PDG} (also termed profile construction) is used to construct confidence intervals, using $q(\sigma)$.
Sensitivities, as well as the coverage properties of the confidence band, detailed in Section~\ref{sec:coverage}, 
are explored with toy-MC simulations. Fig.~\ref{fig:sensi_construction} shows the distributions of lower and upper limits for a background-only simulation, illustrating the sensitivity computation.  Both the science and calibration data-sets 
are drawn from their model distributions, as are the ancillary measurements according to their uncertainties. For the 
sensitivity computation, 1000 toy-MCs are run per mass point. 

Before unblinding, the XENON1T collaboration resolved to use a higher threshold for reporting upper limits than the 1.28\,$\sigma$ corresponding to a $90\%$ confidence level. Only the upper edge of the confidence 
interval is reported until the discovery significance reaches 3\,$\sigma$. This leads to over-coverage at low WIMP cross section values, with an example shown in Fig.~\ref{fig:coverage} affecting limits below the 1\,$\sigma$ sensitivity band. 

The FC construction may produce upper limits excluding signal strengths to which the experiment has a very small discovery power. 
In the case of a downwards fluctuation with respect to the background model, a $15\%$ power-constraint would be applied to set a lower threshold for the upper limit, as proposed in~\cite{PCL}.

The discovery significance is computed using the test-statistic evaluated at the null-hypothesis, 
$q_0 \equiv q(\sigma=0)$. Denoting the distribution of $q_0$ under the null-hypothesis $H_0$ as $f_{H_0}(q_0)$, 
the significance of a given observed test statistic $q^{\mathrm{obs}}_0$ can be expressed with a p-value,

\begin{equation}\label{eqn:discovery}\large
p_{H_0} = \int_{q^{\mathrm{obs}}_0}^\infty\left[f_{H_0}(q_0)\right]\ud q_0. 
\end{equation}
In practice, the p-value is estimated using toy-MC samples to account for non-asymptoticity due to the low signal-background 
overlap. For the FC construction, the test statistic distribution is estimated for 20 steps in the true signal for each mass. For each signal step, 2500 toy-MC simulations, including ancillary measurements are generated and fitted. A threshold curve, constructed by a smooth interpolation between the 90th percentile of the test statistic for each signal size is compared with the final log-likelihood ratio to construct the FC intervals. 

The local discovery significance in Eq.~\ref{eqn:discovery} is computed for a single signal hypothesis. In the spin-independent analysis, the signal hypotheses are the WIMP masses considered between $6~\gevcsq$ and $10000~\gevcsq$. 
To compute a global discovery significance, the distribution of the most significant p-value for any hypothesis is estimated with 10000 toy-MC simulations, and compared with the result from data.

\subsection{The XENON1T Likelihood}
\label{subsec:likelihood}
The log-likelihood used in the spin-independent analysis is a sum of extended un-binned log-likelihoods for the two science runs. Additional terms are the extended un-binned likelihoods for ER calibration data, and terms expressing ancillary measurements of nuisance parameters $\nuis_m$,
\\


\begin{align}\large
    \centering
    \LL_\mathrm{total}(\sigma,\nuiss) =&  \sum_{\sr} \LL_{\sr\ \mathrm{sci}}(\sigma,\nuiss) \nonumber\\
    & +  \sum_{\sr} \LL_{\sr\ \mathrm{cal}}(\nuiss)  \nonumber\\
    & +  \sum_m \LL_m(\nuis_m),
    \label{eqn:totallikelihood}
\end{align}
where $\sr$ runs over data-taking periods, \SRzero and \SRone, $\sigma$ is the WIMP-nucleon cross-section, and $\lagr_\mathrm{sci}$,~$\lagr_\mathrm{cal}$ are the likelihood terms for the WIMP search data and \rnzero calibration data, respectively. Ancillary measurements of nuisance parameters $\nuis_m$ are included in $\LL_m(\nuis_m)$.
The un-binned science likelihood is defined in three dimensions: \cSone, \cStwob and \radius, the radius of the reconstructed event. 
Each background and  signal distributions are defined and normalized in this three-dimensional space. 
The un-binned likelihoods take the form


    \begin{align}\large
        \lagr_{\sr\ \mathrm{sci}}(\sigma,\nuiss) = & \Poiss{N_\sr}{\mu_\mathrm{tot}(\sigma,\nuiss)} \notag\\
        & \times
        \prod_{i=1}^{N_\sr}\left[\sum_c \frac{\mu_c(\sigma,\nuiss) }{\mu_\mathrm{tot}(\sigma,\nuiss)}\cdot f_c(\xmeas_i|\nuiss)  \right], \\
        \mu_\mathrm{tot}(\sigma,\nuiss) \equiv & \sum_c \mu_c(\sigma,\nuiss).
    \label{eqn:unbinned}
    \end{align}
Here, the index $i$ runs over events in the relevant science data-set, and $c$ runs over each signal or background component, with expectation value $\mu_c(\sigma,\nuiss)$, which may be a nuisance parameter or a function of nuisance parameters.
The probability density functions (PDFs), $f_c(\xmeas_i|\nuiss)$, for each component are functions of the analysis coordinates $\xmeas_i = (\mathrm{cS1}_i, \mathrm{cS2}_{b}\,{}_i, \mathrm{R}_i)$ and are evaluated for each event in the likelihood. 
The models for the science data measurements, both the WIMP signal model, and the ER, neutron, \cevns, surface and AC backgrounds are described in detail in the previous sections. 
In the case of the calibration likelihood, which utilizes a smaller volume, 0$<\radiusm<$36.94\,cm, the likelihood uses only the \cSone and \cStwo dimensions, but the  structure of $ \lagr_{\sr\ \mathrm{cal}}$ is otherwise identical to Eq.~\ref{eqn:unbinned}.

The extension of detector modelling to include \radius was made to improve the sensitivity to a 50~\gevcsq~WIMP by $10\%$, 
with respect to an optimized, smaller volume without radial modelling. 
The radial cut is set when the surface background starts dominating and no significant improvement in sensitivity can be obtained.
In addition to the three analysis dimensions in which the distributions are modelled, events in the science data-sets are also classified as inside or outside a core mass of $0.65\ \mathrm{t}$ (shown in Fig.~\ref{fig:ms_neutron}). 
Practically, this can be considered as a combination of two separate un-binned likelihoods in the shape of Eq.~\ref{eqn:unbinned},
%
%
%
%
%
where the relative expectations inside and outside the core mass are determined using the $\radiusm,z$ distribution of the different components. 

\subsection{Nuisance Parameters}

\begin{table*}[htp]
    \centering
    \caption[List of all Likelihood Parameters]{Table of parameters of the full XENON1T likelihood, best-fit values, and $1\,\sigma$ confidence interval for a  $200~\gevcsq$ WIMP. In addition, the data-set that constrains the parameter is noted. Parameters are arranged according to whether they affect background rates, the shape of the background distributions, or the signal distribution. For the shape parameters, the models affected by the parameter are also noted. Due to the mismodelling parameter, the signal distribution, which changes with the WIMP mass, may also affect the ER model best-fit. Expectation values are written as events in each dataset; SR0 (SR1) expectation values refer to a  $0.12~\mathrm{t\times y}$ ($0.88~\mathrm{t\times y}$) exposure. The radiogenic neutron event rate, shared between the science runs, refers to the $1.0~\mathrm{t\times y}$ exposure.
    }
    \label{tab:likelihoodparameters}
        \begin{tabularx}{\linewidth}{XXXr}\toprule
        \hline \hline
        Rate Parameter & & Constraint & Expectation Value\\
        \hline
        \midrule
        \rule{0pt}{1ex}  \\
        \srz~AC events &  & Ancillary measurement & $0.051_{-0 }^{+0.035 }$\\
        \sro~AC events & & Ancillary measurement          &$0.42_{-0 }^{+0.29 }$\\
        \srz~\rnttz~ AC events && Ancillary measurement  &$0.73_{-0 }^{+0.51 }$\\
        \sro~\rnttz~ AC events  && Ancillary measurement  &$2.8_{-0}^{+1.9}$\\
        \srz~\cevns~events & &  Ancillary measurement        &$0.0040_{-0.0014 }^{+0.0014 }$\\
        \sro~\cevns~events && Ancillary measurement        &$0.050_{-0.017}^{+0.017}$\\
        \srz~science data ER && \srz~ data-set        &$73.4_{-8.4}^{+8.8}$\\
        \sro~science data ER && \sro~ data-set        &$554_{-24}^{+24}$\\
        \srz~calibration data ER && \srz~ \rnttz~ data-set    &$689_{-26}^{+27}$\\
        \sro~calibration data ER && \sro~ \rnttz~ data-set    &$5264_{-72}^{+73}$\\
        Radiogenic events in \srz+\sro && Ancillary measurement   &$1.44_{-0.66}^{+0.65}$\\
        \srz~ surface events && \srz~science data           &$12.5_{-3.2}^{+3.9}$\\
        \sro~ surface events && \sro~science data       &$93.7_{-9.5}^{+10}$\\
        \toprule
        \hline\hline
        Shape Parameter &Model Affected& Constraint & Value\\
        \hline
        \midrule
        \rule{0pt}{1ex} \\
        ER photon yield& All ER models& \rnttz~calibration&$-0.043_{-0.052 }^{+0.053 }$\\
        \srz~ER recombination fluctuation &\srz~\rnttz, science models& \srz~\rnttz~data-set        &$0.64_{-0.70}^{+0.58}$\\
        \sro~ER recombination fluctuation&\sro~\rnttz, science models & \sro~\rnttz~data-set        &$0.32_{-0.39}^{+0.41}$\\
        \srz~safeguard as fraction of ER&\srz~\rnttz, science models & \srz~\rnttz~data-set   &$-0.0060_{-0.0036}^{+0.0049}$\\
        \sro~safeguard as fraction of ER&\sro~\rnttz, science models & \sro~\rnttz~data-set       &$-0.0049_{-0.0045}^{+0.0014}$\\
        \srz~surface shape parameter &\srz~surface model& \srz~science data&$1.00_{-0.67}^{+0.0}$ \\
        \sro~surface shape parameter&\srz~surface model & \sro~science data&$-0.39_{-0.22}^{+0.24}$\\
        \toprule
        \hline\hline
        Signal Parameter& & Constraint &Value\\
        \hline
        \midrule
        \rule{0pt}{1ex} \\
        \srz~ signal efficiency && NR model uncertainty             &$1.000_{-0.036}^{+0.036 }$\\
        \sro~ signal efficiency && NR model uncertainty             &$1.000_{-0.021}^{+0.021 }$\\
        WIMP cross-section [$10^{-45}\mathrm{cm}^2$]&& &$0.0421_{-0}^{+0.1165}$ \\
        WIMP mass [GeV/c$^2$] &Signal, ER mismodelling term& Fixed in analysis & 200\\
        \bottomrule
        \hline
    \end{tabularx}
\end{table*}

The signal and background models, consisting of expectation values and distributions in analysis space, depend on several nuisance parameters. Table~\ref{tab:likelihoodparameters} lists all the parameters of the combined likelihood, the data sets that mainly constrain them and their best-fit value. Uncertainties on the nuisance parameters are computed using the profiled likelihood in each nuisance parameter, but unlike the signal confidence intervals, the asymptotic construction is applied. Nuisance parameters are grouped in rate parameters, the expectation value for all background components, signal efficiency parameters and shape parameters that affect the model distributions. Expectation values for the 5 modelled backgrounds in the science data, as well as AC and ER rates in the \rnzero calibration likelihood, are all nuisance parameters in the likelihood. In the following, a separate nuisance parameter and term is applied for each science run, with the exception of the radiogenic rate.

The radiogenic rate, as well as the expected CE$\nu$NS rates for each science run, is  constrained by ancillary measurements expressed as Gaussian likelihoods,

\begin{equation}\large
    \lagr_m(\mu_c) =  \Gauss{\hat{\mu}_c}{\mu_c}{\sigma_{c}}.
\end{equation}
These likelihoods are defined by the PDF for the ancillary measurement of the component $\hat{\mu_c}$, given a true expectation value $\mu_c$ and measurement uncertainty $\sigma_c$. 
The signal expectation is 
\begin{equation}\large
\mu_{\mathrm{sig}} = \sigma \cdot \epsilon \cdot \mu_{\mathrm{ref}}(M)/\sigma_\mathrm{ref}
\end{equation}
where $\mu_{\mathrm{ref}}$ 
is the reference expectation for a WIMP of mass $M $ and cross-section $\sigma_\mathrm{ref}$, given by the signal acceptance, and $\epsilon$ is a multiplicative factor expressing the uncertainty on the signal expectation for a fixed cross-section.
This expectation uncertainty is constrained by the NR model posterior taking all model variations into account,

\begin{equation}\large
    \lagr_\mathrm{m}(\epsilon) = \Gauss{1}{\epsilon}{\sigma_{\epsilon}(M)},
\end{equation}
where the uncertainty $\sigma_{\epsilon}(M)$ depends on the WIMP mass, ranging from $0.15$ at $6~\gevcsq $ to $0.03$ at $200~\gevcsq$, as shown in the inset of Fig.~\ref{fig:wimp_signal_model}.
The nominal value of $1$ reflects that the best-fit expectation is expressed by the reference expectation. 

The AC rate $\mu_\mathrm{AC}$ is constrained between two extreme estimates of its rate and is assigned a uniform PDF between the lower and upper reference, written as $0.6\cdot\hat{\mu_\mathrm{AC}}$ and $\hat{\mu_\mathrm{AC}}$ for convenience,

\begin{equation}\large
    \lagr_\mathrm{m}(\mu_{\mathrm{AC}}) =  \Uniform{\mu_{\mathrm{AC}} }{0.6\cdot \hat{\mu}_{\mathrm{AC}}}{\hat{\mu}_{\mathrm{AC}}}.
\end{equation}
The ER and surface background rates are not assigned auxiliary measurements, as their high statistics in the science data sample constrain them. In the case of the surface background, the region of highest signal overlap (at low \cSone and \radius) was blinded. This motivated the conservative procedure of not placing an auxiliary constraint on the surface shape. 

In addition to uncertainties on the rate, 
the ER and surface background PDFs are also assigned shape uncertainties. 
The surface background sideband fit, described in Section~\ref{sec:surface}, is also used to construct an uncertainty for the radial slope of the background. 
To avoid over-constraining this distribution, which could lead to spurious excesses or too-tight confidence intervals, no constraint is placed on the radial slope from the sideband measurement. The science data fit finally constrains the radial slope with approximately $2.5$ times smaller uncertainty than the sideband fit. 

The ER model is described in detail in Section~\ref{subsec:ER}. The nuisance parameters $\gamma_\mathrm{er}$ and $\Delta r$ described there are propagated to the likelihood, with the former shared between science runs. 
%
The value of these nuisance parameters and their associated errors are determined in the combined fit by including the ER calibration likelihood term for \rnzero data. This ER model includes the nominal ER model, as well as variations due to changing the photon yield or recombination fluctuation parameter. It is slightly different from the one illustrated in Section~\ref{sec:detector_signal_response_model}, as the uncertainties in the model from nuisance parameters other than $\gamma_\mathrm{er}$ and $\Delta r$ are not included in the nominal ER model.

\begin{figure}[tbp]
\centering
\includegraphics[width=\columnwidth]{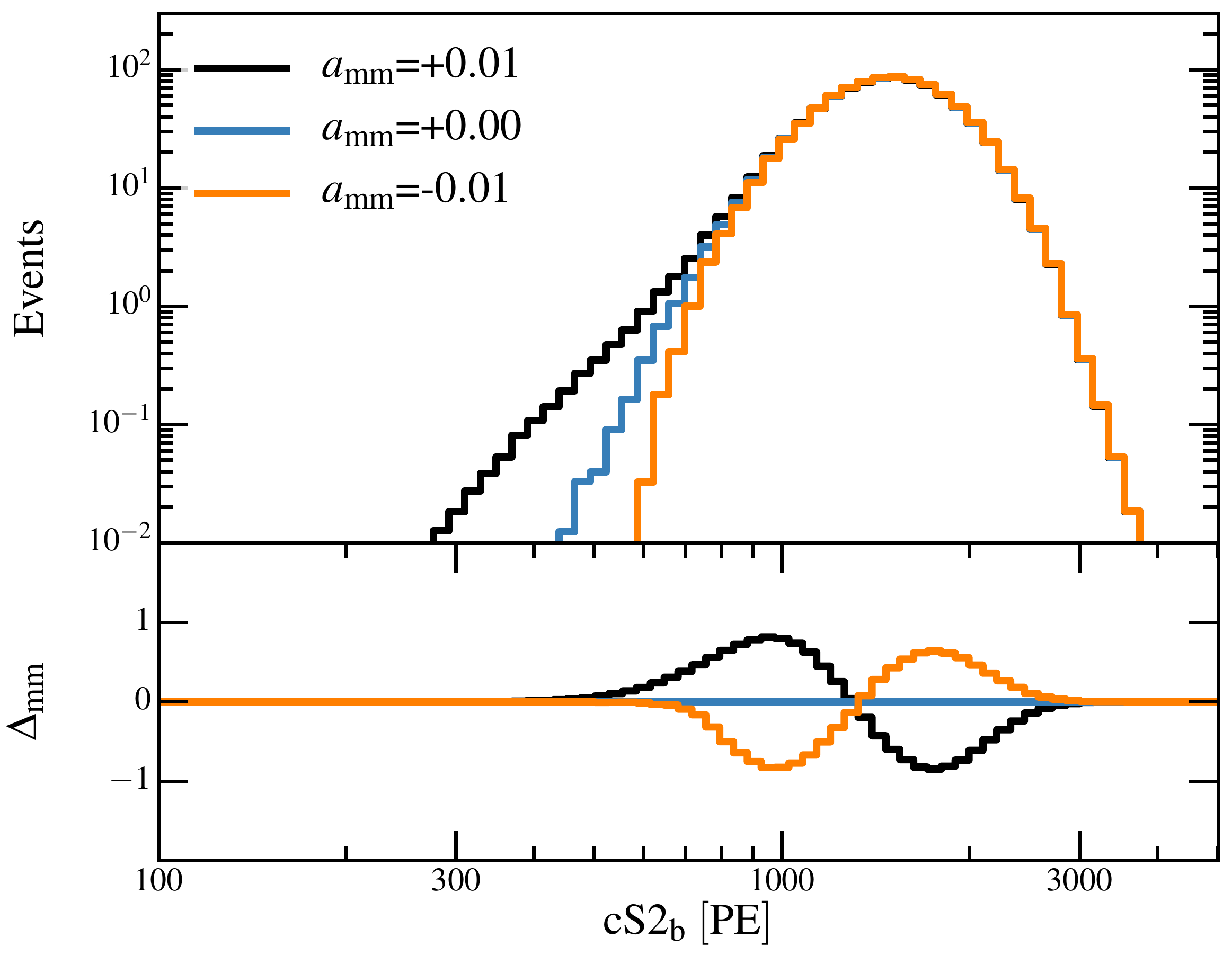}
\caption{\label{fig:safeguard} 
Illustration of the effect of a mismodelling term on the ER model for a slice of parameter space $20<\cSone<30$\,PE, 
projected onto cS2$_b$. The upper curve shows the ER model for safeguard fractions $-0.01$, $0$ and $0.01$, showing how a positive (negative) mismodelling term raises (lowers) the signal-like tail of the ER model with respect to the nominal model in blue. 
The lower panel shows $\Delta_\mathrm{mm}$, the difference between each model and the nominal, demonstrating both the effect on the tail at low cS2$_b$ and the opposite at higher cS2$_b$, due to the normalization of the PDF.
}
\end{figure}

A mismodelling term, or "safeguard", proposed in~\cite{safeguard}, is a shape-uncertainty added to the ER model, consisting of a signal-like component added to or subtracted from the nominal ER model. This ensures that regardless of other nuisance parameters, the ER model will have the freedom to fit the calibration data in the signal-like region. A spurious signal-like over- or under-estimation in the ER background will have the greatest impact on the inference, giving spuriously constraining limits or spuriously significant excesses, respectively. The safeguard mainly affects the ER model tail which overlaps with the signal region, as shown in Fig.~\ref{fig:safeguard}. 
The ER model, including the mismodelling term is constrained by the calibration likelihood terms included in the total likelihood, this allows the safeguard component 
to be constrained by the much higher (approximately 10 times for \SRzero and \SRone)  statistics of the \rnzero calibration data compared to the science data. 
The  ER background PDF including the safeguard term,  PDF $f^\prime_\mathrm{ER}(\xmeas| \nuiss_\mathrm{ER})$ can be written as

%

\begin{equation}\large
        f^\prime_\mathrm{ER} \equiv  \bmm\cdot\left[ (1-\amm)\cdot f_\mathrm{ER}(\xmeas| \nuiss_\mathrm{ER}) + \amm \cdot f_\mathrm{sig}(\xmeas)\right],
\end{equation}
where $\amm$ is the safeguard nuisance parameter and $f_\mathrm{sig}$ the signal PDF. If the safeguard causes the PDF to be negative in a region, it is truncated to 0. The pre-factor $\bmm$, a function of $\amm$ and nuisance parameters that affect the ER distribution $\nuiss_\mathrm{ER}$, ensures that the PDF is normalized in the analysis space. 

The compatibility between the best-fit to data and the safeguard-equal-zero hypothesis is performed using the profiled log-likelihood ratio to compute the 1-sigma error on the best-fit safeguard. The calibration data indicates that the signal-like tail of the ER is less pronounced than in the nominal model, reflected in a negative safeguard fit. A zero safeguard is $2.9\,\sigma$ from the \SRone+\SRzero combined fit using this method. 

\subsection{Coverage}
\label{sec:coverage}

\begin{figure}[htbp]
\centering
\includegraphics[width=\columnwidth]{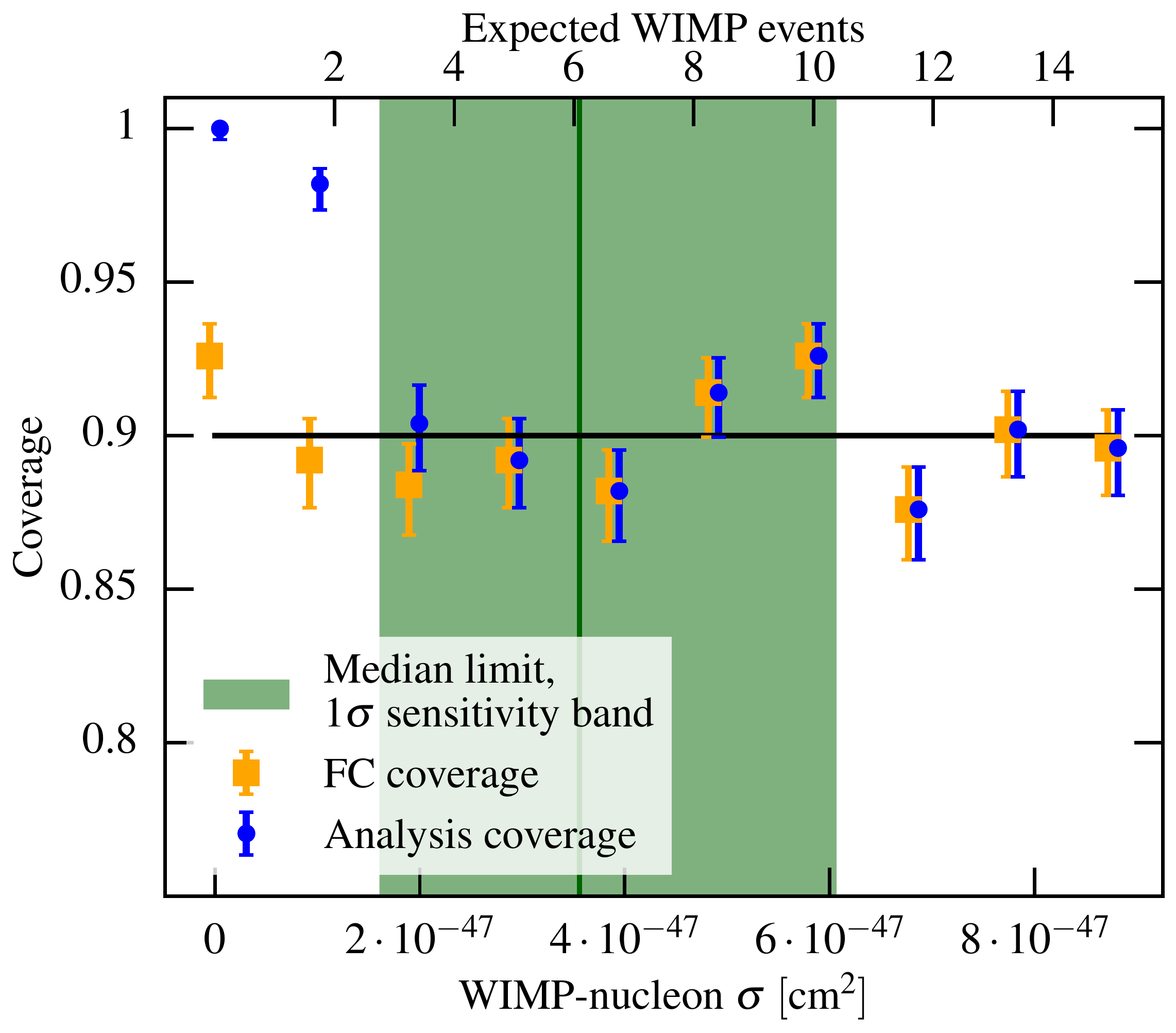}
\caption{\label{fig:coverage} Figure displaying the coverage of confidence intervals for the spin-independent 1 tonne-year analysis and a 50 \gevcsq~WIMP. Error-bars indicate $1\,\sigma$ confidence intervals around the best estimate. Orange squares  show the result using the profile construction, while the blue circles show the coverage of the XENON1T analysis including the $3\,\sigma$ threshold for reporting upper limits. The upper and lower x-axes show the WIMP cross-section and expectation (setting $\epsilon=1$) respectively, and the green band and line highlight the sensitivity and $1\,sigma$ band of upper limits for the analysis. }
\end{figure}

The fraction of repeated experiments where the confidence interval contains the true parameter is called the coverage. Perfect coverage is equal to the confidence interval, $0.9$ in the case of XENON1T. While the nominal FC construction, introduced in section~\ref{inferencemethod}, provides coverage by construction, the coverage of the profile construction must be investigated for the likelihood in question. As we decided to report only the upper edge of the confidence interval for discovery significances $<3\,\sigma$, there will be over-coverage at very low signal sizes. This has a similar effect to the power constraint. Fig.~\ref{fig:coverage} shows the coverage for a $50$ \gevcsq~WIMP, both for the profile construction (orange points), and the analysis including the $3\,\sigma$ threshold (blue points). The green band shows the $-1\,\sigma$ to $1\,\sigma$ sensitivity band. The result is consistent with perfect coverage (black line), with overcoverage for the $3\,\sigma$ threshold only under the $-1\,\sigma$ edge of the sensitivity band. 
The effect on coverage of mismeasuring the nuisance parameters was also studied. In the case of the mismodelling term, a shift in the parameter by three times the observed value was required for a one percentage point shift in the coverage.

\section{Summary}
\label{sec:summary_outlook}
In this manuscript we have reported  details of the detector response model, the background and WIMP signal models, and the statistical inference of XENON1T analysis chain.
These have been used in the interpretation of results in the search for spin-independent elastic WIMP-nucleon interactions in XENON1T~\cite{xe1t_sr0, xe1t_sr1}, and is also used in searches for alternative dark matter candidates or interactions using XENON1T data~\cite{xenon_1t_pion,xenon_1t_sd}.
The background model, using  a simulation-based detector response model, has been improved with respect to previous analyses by reduction of the systematic bias and a better treatment of parameter uncertainties and correlations.
The statistical inference was developed to use a combined, unbinned likelihood that is adaptable to multiple analysis spaces and additional data sets. A signal-like mismodeling term is introduced for the first time in an analysis as a background model shape uncertainty. In addition, the inference now employs toy Monte Carlo simulations extensively to construct and validate the confidence bands.

\section*{Acknowledgement}
We gratefully acknowledge support from the National Science Foundation, Swiss National Science Foundation, German Ministry for Education and Research, Max Planck Gesellschaft, Deutsche Forschungsgemeinschaft, Netherlands Organisation for Scientific Research (NWO), Netherlands eScience Center (NLeSC) with the support of the SURF Cooperative, Weizmann Institute of Science, Israeli Centers Of Research Excellence (I-CORE), Pazy-Vatat, Initial Training Network Invisibles (Marie Curie Actions, PITNGA-2011-289442), Fundacao para a Ciencia e a Tecnologia, Region des Pays de la Loire, Knut and Alice Wallenberg Foundation, Kavli Foundation, and Istituto Nazionale di Fisica Nucleare. Data processing is performed using infrastructures from the Open Science Grid and European Grid Initiative. We are grateful to Laboratori Nazionali del Gran Sasso for hosting and supporting the XENON project.


\end{document}